\pdfoutput=1

\documentclass[preprint]{aastex6}

\usepackage{amsmath}
\usepackage{bm}
\usepackage{tabularx}

\fullcollaborationName{The Friends of AASTeX Collaboration}

\begin{document}


\title{Mean motion resonances at high eccentricities:\\ the 2:1 and the 3:2 interior resonances}


\author{Xianyu Wang\altaffilmark{1} and Renu Malhotra\altaffilmark{2}}
\affil{The University of Arizona}


\altaffiltext{1}{Also Tsinghua University, xianyu@lpl.arizona.edu}
\altaffiltext{2}{renu@lpl.arizona.edu}

\begin{abstract}

Mean motion resonances [MMRs] play an important role in the formation and evolution of planetary systems and have significantly influenced the orbital properties and distribution of planets and minor planets in the solar system as well as exo-planetary systems. Most previous theoretical analyses have focused on the low-to-moderate eccentricity regime, but with new discoveries of high eccentricity resonant minor planets and even exoplanets, there is increasing motivation to examine MMRs in the high eccentricity regime. Here we report on a study of the high eccentricity regime of MMRs in the circular planar restricted three-body problem. Non-perturbative numerical analyses of the 2:1 and the 3:2 interior resonances are carried out for a wide range of secondary-to-primary mass ratio, and for a wide range of eccentricity of the test particle. The surface-of-section technique is used to study the phase space structure near resonances. We identify transitions in phase space at certain critical eccentricities related to the geometry of resonant orbits; new stable libration zones appear at high eccentricity at libration centers shifted from those at low eccentricities. We present novel results on the mass and eccentricity dependence of the resonance libration centers and their widths in semi-major axis.  Our results show that MMRs have sizable libration zones at high eccentricities, comparable to those at lower eccentricities.

\end{abstract}


\section{Introduction} \label{sec:intro}

There are many examples of stable mean motion resonances (MMRs) in our solar system. The three inner Galilean moons of Jupiter -- Io, Europa, and Ganymede -- are in pair-wise 2:1 mean motion resonances, and together their orbital motions are locked in the chain of mean motion resonances, 4:2:1, known as the ``Laplace Resonance"; these resonances are stable under tidal evolution~\citep{Peale76,Peale99}. Several pairs of moons of Saturn are also locked in MMRs, such as Mimas and Tethys in a 2:1, Enceladus and Dione in a 2:1, and Titan and Hyperion in a 4:3 MMR ({\it ibid.}).  The Jupiter Trojans are a large group of asteroids orbiting around Jupiter's two stable Lagrangian points, L4 and L5; these are locked in the 1:1 stable orbital resonances with Jupiter \citep{Levison97}.  Pluto is locked in a 2:3 mean motion resonance with Neptune; Pluto's perihelion distance of $29.7$ AU is less than the semimajor axis of Neptune $30.1$ AU, so it travels in a Neptune-crossing orbit; but the stable 2:3 MMR with Neptune keeps Pluto from colliding with Neptune \citep{Malhotra:1997}.  Many Kuiper belt objects are found in the 2:3 MMR and other MMRs with Neptune.  Planetary mean motion resonances may also exist in exoplanetary systems \citep{Marcy:2001,Lee:2002,Fabrycky:2014,Petigura15,Nelson16,Mills:2016}.  Furthermore, unstable mean motion resonances are also of great significance; for example, the Kirkwood Gaps in the asteroid belt are linked to the unstable and chaotic MMRs with Jupiter \citep{Moons:1996}, and the stability of our planetary system itself is intimately related to the role of MMRs \citep{Murray:1999}

The phase space structure near MMRs is quite complex, and there is a considerable literature on their mathematical analysis.  The planetary three body problem, with $m_1\gg m_2,m_3$, is the usual starting point of such studies.
Although it does not describe the full complexities of MMRs in the real solar system, it is still a very useful approximation for low-eccentricity and low-inclination planetary systems and provides important insights for the phase space structure of MMRs.
The special case of the circular planar restricted three body problem is particularly important in studies of the dynamics of small bodies in planetary systems. In this special case, two massive bodies, $m_1\gg m_2$, move in circular orbits about their center of mass, and a massless third body (the test particle, $m_3=0$) moves in the massive bodies' orbital plane.  This problem admits an integral of the particle's motion, the Jacobi integral $C_J = 2(\Omega L-E)$, where $\Omega$ is the angular velocity of the massive bodies in their circular orbits and $E$ and $L$ are the specific energy and specific angular momentum of the particle.  Theoretical analysis of MMRs, based on perturbation theory, in this special case can be found in \citet{Murray99}. In particular, we mention the analytical results for the widths of the interior MMRs of Jupiter for low to moderate eccentricities of the test particle \citep[sect.~8.8]{Murray99}.
Semi-analytical perturbative methods have been used to map the widths of stable resonance libration zones for a wider range of eccentricity, for the major exterior MMRs of Neptune \citep{Morbidelli:1995}; however, in this case, non-perturbative numerical analysis shows that the semi-analytical results overestimate the stable resonance zones and give poor accuracy at eccentricities exceeding $\sim0.2$ \citep{Malhotra96}.
Meanwhile, recent observations find that high eccentricity MMRs may not be uncommon in our solar system. \citet{Michel:2000} noted the importance of MMRs in the dynamics of planet-crossing asteroids in the inner solar system. \citet{Chiang:2003} noted the surprising existence of high eccentricity, $e\gtrsim0.4$, Kuiper belt objects in the exterior 2:5 MMR of Neptune. \citet{Malhotra16} suggest that some of the most distant Kuiper belt objects (Sedna, 2010 GB174, 2004 VN112, 2012 VP113), with eccentricities in excess of $\sim0.7$, may be locked in interior MMRs with an unseen distant planet.

Our goal is to understand how the widths of stable libration zones of MMRs behave at very high eccentricities.  The problem is sufficiently challenging that we limit the present work to only two important cases, the interior 2:1 and 3:2 MMRs in the circular planar restricted three body model.
We use non-perturbative numerical analysis to calculate the resonance widths for test particle eccentricities as large as 0.99. The resonance libration regions are visualized in surfaces of section. Contrary to the common expectation that resonance overlap leads to chaos and loss of stable resonant librations at high eccentricities, we report that large new stable libration zones reappear at higher eccentricity, albeit at shifted libration centers.  We report several novel results, including the transitions of the resonance libration centers with increasing eccentricity, the variation of resonance width with eccentricity, and its dependence on the mass ratio $\mu=m_2/(m_1+m_2)$.  Our results would be of interest in a wide range of astronomical applications of the restricted three body problem.

\section{Surfaces of Section} \label{sec:sos}

In the planar circular restricted three body problem, we will refer to the most massive body, of mass $m_1$, as the ``sun", and the secondary mass, $m_2,$ as the ``planet", and the third (massless) body as the ``test particle".  All the bodies are restricted to move in a common plane, the $(x,y)$ plane. We adopt the natural units for this problem: the unit of length is the orbital separation of $m_1$ and $m_2$, the unit of time is their orbital period divided by $(2\pi)$, and the unit of mass is $m_1+m_2 \simeq m_1$; with these units the constant of gravitation is unity, and the orbital angular velocity of $m_1$ and $m_2$ about their common center of mass is also unity.
Then, in a rotating reference frame, of constant unit angular velocity and origin at the center of mass, both $m_1$ and $m_2$ remain at fixed positions, $ (-\mu,0) $ and $ (1-\mu,0) $, respectively, where $ \mu=m_2/(m_1+m_2) $. The equations of motion of the test particle can be written as
\begin{equation}\label{eq5}
\ddot{x}=2\dot{y}+x-\frac{(1-\mu)(x+\mu)}{{r_1}^3}-\frac{\mu(x-1+\mu)}{{r_2}^3},
\end{equation}
\begin{equation}\label{eq6}
\ddot{y}=-2\dot{x}+y-\frac{(1-\mu)y}{{r_1}^3}-\frac{\mu y}{{r_2}^3}.
\end{equation}
where $ r_1 $ and $ r_2 $ are the test particle's distances to the sun and the planet, respectively,
\begin{equation}\label{eq3}
r_1=\left[ (x+\mu)^2+y^2 \right]^{1/2},
\end{equation}
\begin{equation}\label{eq4}
r_2=\left[ (x-1+\mu)^2+y^2 \right]^{1/2}.
\end{equation}
The Jacobi constant can be derived from the dynamic equations as
\begin{equation}\label{eq7}
C_J=x^2+y^2-\dot{x}^2-\dot{y}^2+\frac{2(1-\mu)}{r_1}+\frac{2\mu}{r_2},
\end{equation}
which can also be expressed in terms of orbital elements as
\begin{equation}\label{eq8}
C_J = \frac{1}{a}+2\sqrt{a(1-e^2)(1-\mu)}+O(\mu),
\end{equation}
where $ a $ and $ e $ are semimajor-axis and eccentricity of the particle's osculating orbit about the sun.

\begin{figure}[ht!]
\figurenum{1}
\plotone{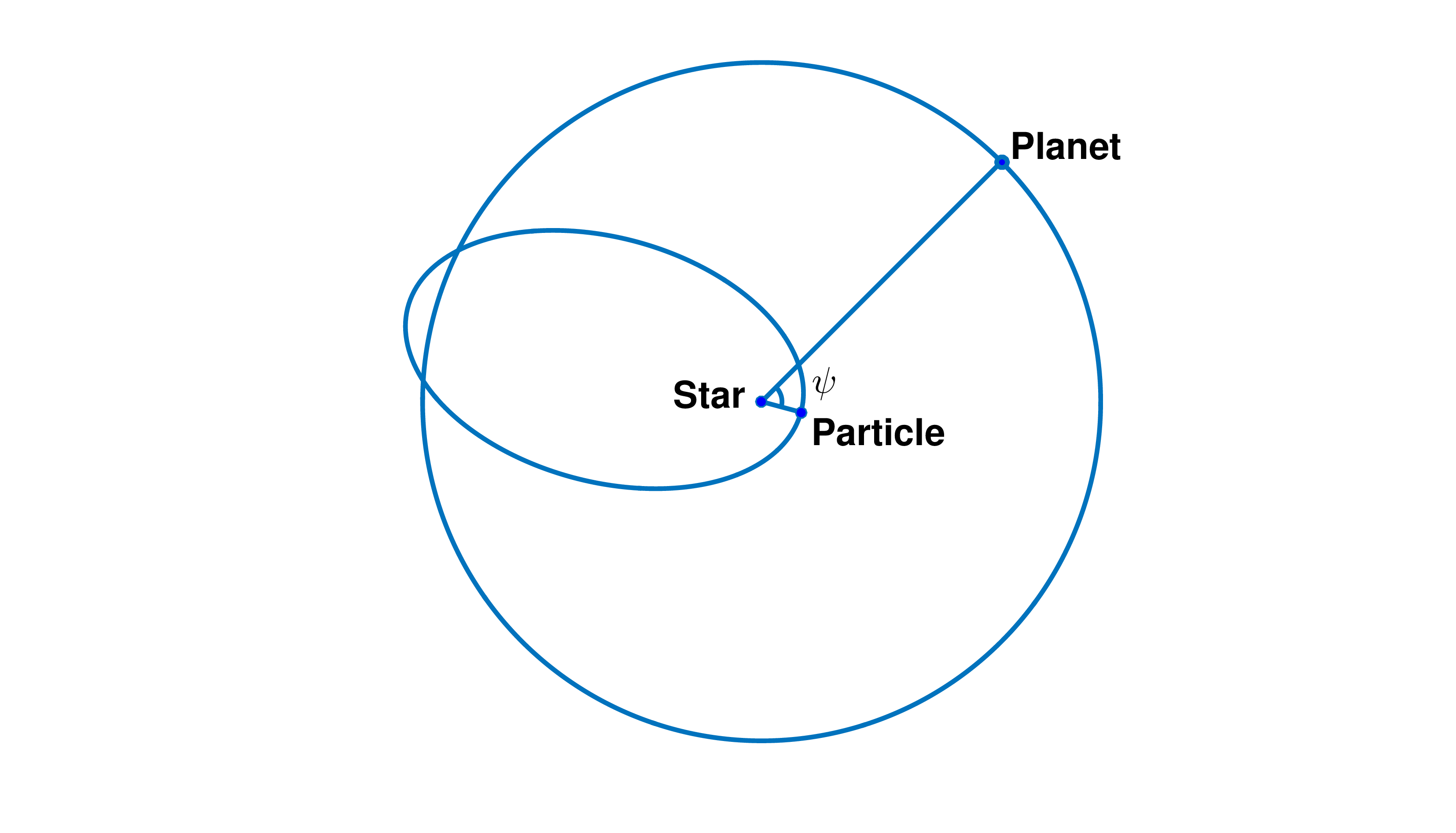}
\caption{Illustration for the surface of section: $\psi$ is the angular separation of the planet and test particle's pericenter.
\label{fig:f1}}
\end{figure}

The motion of the test particle takes place in the four-dimensional phase space $(x, y, \dot{x}, \dot{y})$. With the constraint of the Jacobi constant, the test particle's motion is constrained to a three dimensional surface in this phase space.  To visualize the phase space structure in a two-dimensional space, we compute ``surfaces of section" near the MMRs of interest. A surface of section is analogous to the stroboscopic plot of the phase space variables of a periodically forced nonlinear oscillator \citep{Lichtenberg:1983}.   Our implementation for the two-degrees of freedom of our test particle dynamics is as follows.
On a continuous phase space trajectory of the test particle with specified initial conditions (and with a specified value of the Jacobi constant), we identify its intersections with the surface of section with the following condition,
\begin{equation}\label{eq9}
\dot{r}_1=0,\ddot{r}_1>0.
\end{equation}
That is, the phase space trajectory of the particle is represented by the collection of its phase space variables $(x,\dot x,y,\dot y)$ recorded at its pericenter passages.  The phase space structure near the MMR is obtained by computing this representation for many test particles with different initial conditions in the neighborhood of the MMR, but having the same value of the Jacobi constant.  However, rather than making the usual two-dimensional plots of $(x,\dot x)$ or $(y,\dot y)$, we transform the phase space variables to osculating orbital elements and make plots of the orbital parameters $ (a, \psi) $, where $ a $ is the test particle's semi-major axis and $ \psi $ is the angular separation of the planet and the particle at periapsis, as illustrated in Figure~\ref{fig:f1}.

Successive points on the surface of section that are confined to smooth curves in the $ (a, \psi) $ plane indicate a quasi-periodic stable orbit of the test particle, whereas successive points filling an area in this plane indicate a chaotic orbit. A closed curve on which $\dot\psi$ changes sign indicates an orbit in resonant libration. The stable resonant zones (and their widths in semi-major axis) are then readily visualized in such surfaces of section.

The detailed procedure for producing the surfaces of section is as follows.
First, for a specified initial value of $a$ and $e$, we calculate the Jacobi constant $C_J$ using Eq.\eqref{eq8}. (This first value of $a$ is always chosen as the resonant value, $a_{res}$, see Eq.~\ref{e:ares} below.)  Next, with a specified initial value of $\psi$, we calculate the initial position and velocity vectors $(x_p,y_p)$, $(\dot{x}_p,\dot{y}_p)$ of the test particle at pericenter in the rotating frame:
\begin{equation}\label{eq10}
(x_p,y_p)=r_p(\cos(\psi)-\mu,-\sin(\psi)),
\end{equation}
\begin{equation}\label{eq11}
(\dot{x}_p,\dot{y}_p)=v_p(\sin(\psi),\cos(\psi)),
\end{equation}
where $r_p=a(1-e)$ and $v_p=(\dot{x}_p^2+\dot{y}_p^2)^{1/2}$ is calculated with the help of Eq.\eqref{eq7},
\begin{equation}
v_p=\sqrt{ x_p^2+y_p^2+\frac{2(1-\mu)}{r_1}+\frac{2\mu}{r_2}-C_J}.
\end{equation}
With these initial values of the position and velocity vectors, we numerically integrate Eq.\eqref{eq5}-\eqref{eq6} to obtain the phase space trajectory of the test particle.
Typical integration times run to 60,000 orbital periods of the test particle, equivalent to approximately $188,500$ units of time for 2:1 MMR and $251,300$ units of time for 3:2 MMR. During the integration, we identify each pericenter passage occurrence (by testing for the condition Eq.\eqref{eq9}), and compute the values of $a$ and $\psi$ at each pericenter passage,
\begin{equation}\label{eq12}
\psi=\arctan(y_p, x_p+\mu),
\end{equation}
\begin{equation}\label{eq13}
a=-\frac{1}{2E},
\end{equation}
where $E$ is the orbital energy of the test particle,
\begin{equation}\label{eq14}
E= \frac{1}{2}(\dot{X}^2+\dot{Y}^2)
-\frac{1-\mu}{r_1}-\frac{\mu}{r_2},
\end{equation}
where [$\dot{X}$, $\dot{Y}$] is the velocity vector in the barycentric inertial frame.
For additional trajectories in the same surface of section, we keep the same value of $C_J$ but use new initial values of and $\psi$ and $a$ (close to $a_{res}$), and compute $e$ according to Eq.\eqref{eq8}; with these, we compute new initial values of $(x_p,y_p)$ and $(\dot{x}_p,\dot{y}_p)$. We repeat this procedure several times to fill in the surface of section in the MMR neighborhood.

The neighborhood of an MMR is usually identified as a small range of semi major axis near the exact resonant value in which a critical resonant argument can librate.  A $ j:k $ mean motion resonance occurs when the test particle completes $j$ orbits about the sun while the planet completes $k$ orbits, where $j$ and $k$ are mutually prime numbers. The resonant value of the semi-major axis of the test particle is
\begin{equation}\label{e:ares}
a_{res}=\left(\frac{k}{j}\right)^{2/3}.
\end{equation}
When the eccentricity $e$ is also given, the Jacobi constant for an eccentric resonant orbit can be calculated using Eq.\eqref{eq8} with $a_{res}$ and $e$.  Thus, the surfaces of section in the neighborhood of an MMR can be characterized by either their fixed value of the Jacobi constant or by the combination $(a_{res},e)$.  In the following section, we will label each different surface of section with simply the specific value $e$, since $a_{res}$ is also fixed for each resonance.

The critical resonant argument (in the circular planar restricted three body problem) is defined as
\begin{equation}\label{eq18}
\phi=j\lambda'-k\lambda-(j-k)\varpi,
\end{equation}
where $\lambda$ and $\lambda^{\prime}$ are mean longitudes of the test particle and the planet, and $\varpi$ is the longitude of pericenter of the test particle.
In a stable resonance zone, the resonant argument will librate about a central value, either 0 or $180^\circ$.
When the test particle is at its pericenter, $\lambda=\varpi$, then we have
\begin{equation}\label{eq19}
\phi=j(\lambda^{\prime}-\varpi).
\end{equation}
We can see that the resonant argument is $j$ times the angular element $\psi$ that we defined for the surface of section.

In the present study, we focus on the interior 2:1 and 3:2 MMRs at high eccentricities, $0.05\leq e \leq0.99$, and for values of mass ratio $\mu$ ranging from $ 1 \times 10^{-5} $ to $ 3 \times 10^{-3} $.  For the phase space structure at very small eccentricities, $e\lesssim0.1$, we refer the reader to the textbook by \cite{Murray99}.

\section{Mean Motion Resonance Zones} \label{sec:mmr}

\subsection{Interior 2:1 MMR\label{subsec:2to1}}

A test particle near the 2:1 MMR with the planet has a semi-major axis value close to the resonant value,
\begin{equation}\label{eq20}
a_{res}=\left(\frac{1}{2}\right)^{2/3}=0.630.
\end{equation}
For an elliptic orbit, the maximum aphelion distance is
\begin{equation}\label{eq21}
\lim_{e \to 1} a_{res}\left(1+e\right)=1.260.
\end{equation}
This means that for large eccentricities, $e > a_{res}^{-1}-1 \simeq 0.59$, the test particle's orbit will cross the orbit of the planet.

\begin{figure}
\figurenum{2}
\gridline{\fig{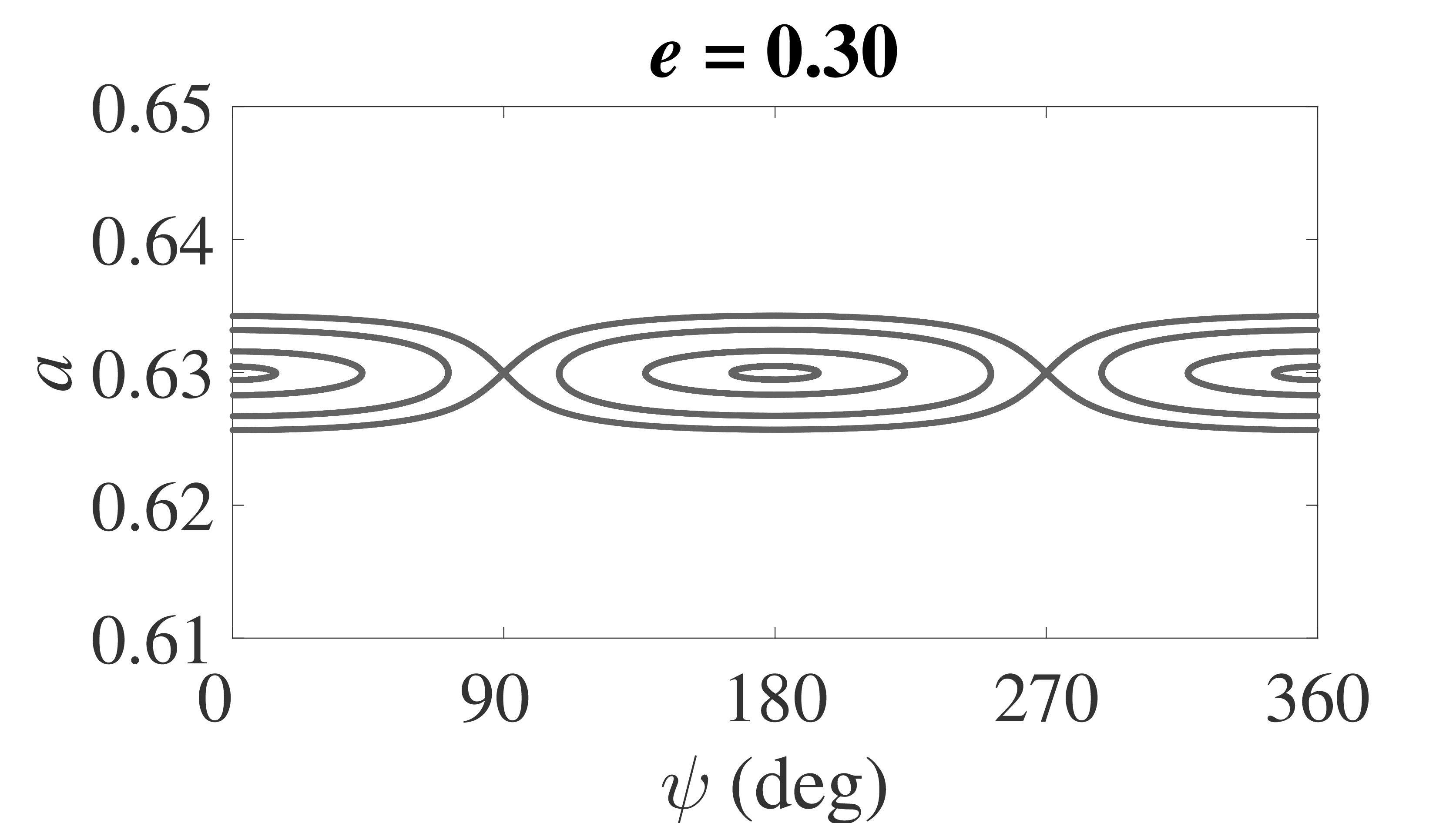}{0.45\textwidth}{(a)}
          \fig{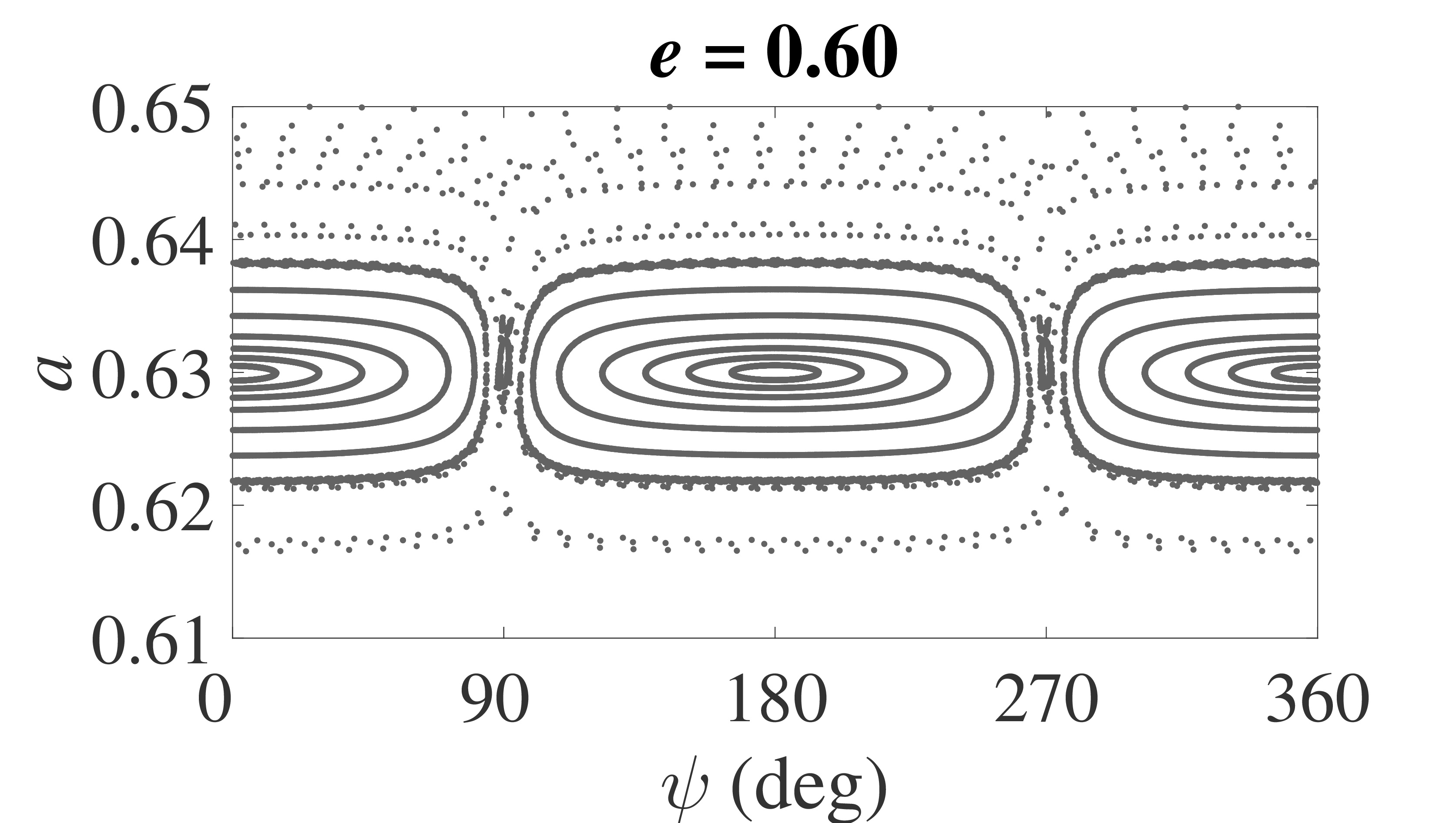}{0.45\textwidth}{(b)}
          }
\gridline{\fig{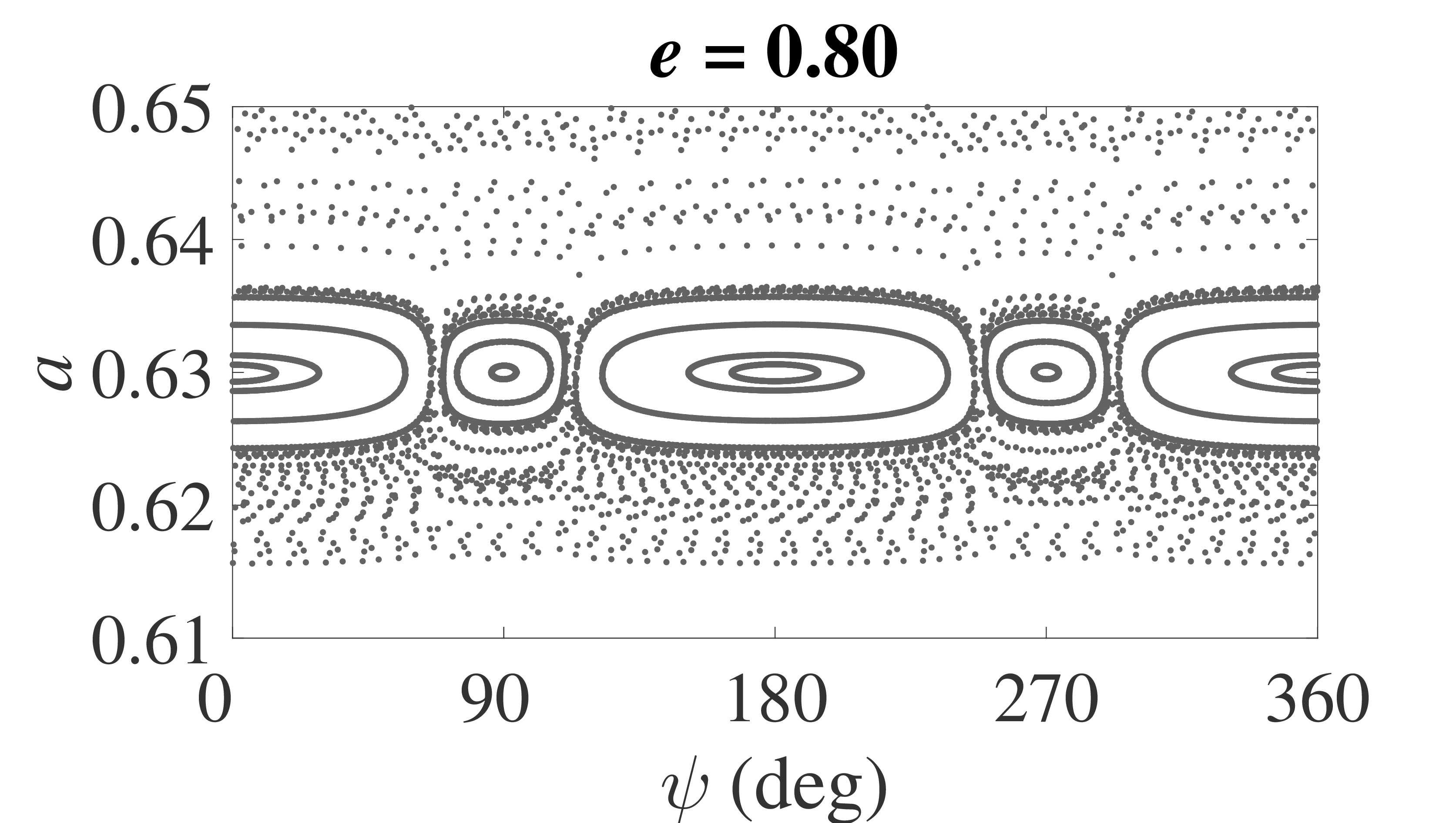}{0.45\textwidth}{(c)}
          \fig{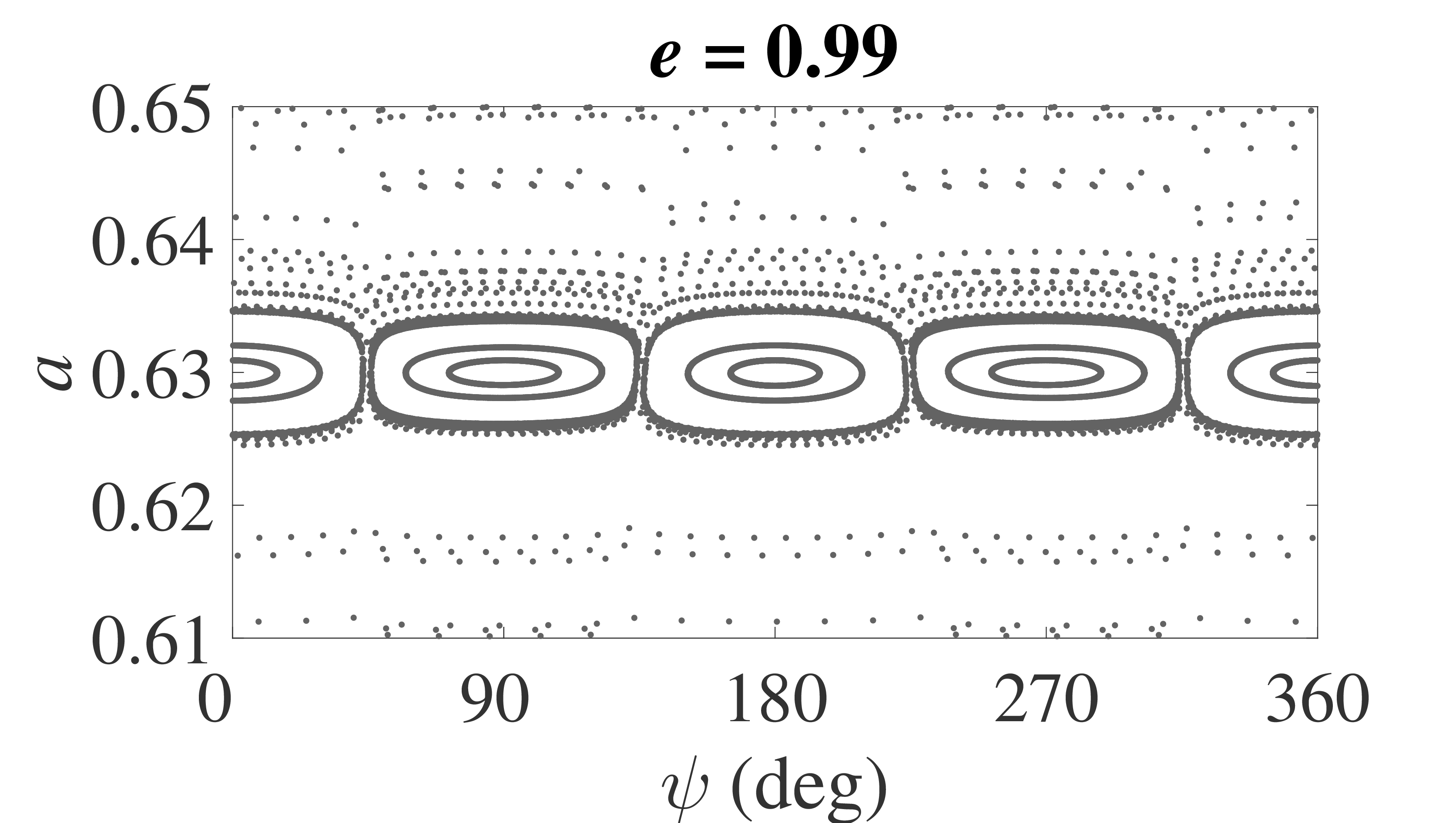}{0.45\textwidth}{(d)}
          }
\caption{Surfaces of section in $(a,\psi)$ near the interior 2:1 MMR, for mass ratio $\mu = 3 \times 10^{-5}$.
Each panel has a fixed value of the Jacobi constant, parametrized by the combination of the resonance value of the semi major axis, $a_{res}=0.630$, and the indicated value of the eccentricity.  The chains of islands in each surface of section indicate the stable libration regions, and the centers of the islands represent the stable periodic orbit in exact
2:1 resonance with the planet.  For $e=0.30$, there is a chain of two islands centered at $\psi=0$ and $\psi=180^\circ$.  For the larger eccentricity values, there is an additional chain of two islands, centered at $\psi=90^\circ$ and $\psi=270^\circ$.
\label{fig:f2}}
\end{figure}

We generated surfaces of section for initial values of the particle's semi-major axis $a$ around the resonant $a_{res}$, various initial values of its angular separation from the planet, $ \psi $, and for a range of initial eccentricities.  Results are shown in Figure \ref{fig:f2} for one value of the mass ratio, $\mu =3 \times 10^{-5}$.
We observe that when the eccentricity is small (Figure \ref{fig:f2}(a)), there are two stable islands in the surface of section, which are centered at $\psi=0^\circ$ and $\psi=180^\circ$;
the boundaries of these stable islands are smooth and no significant chaotic regions exist.
At larger eccentricity, the widths (measured as the range $\Delta a$ where $\psi$ librates) of the stable islands grow; we observe the presence of chaotic regions near their boundaries, and these chaotic regions expand with increasing eccentricity.
When the eccentricity is larger than a certain critical value, $e_{c}\simeq 0.59$, we observe that new stable islands appear in the surface of section, and the number of the stable islands doubles from two to four (Figure \ref{fig:f2}(b)). The new islands are centered at $\psi=90^\circ$ and $\psi=270^\circ$, in-between the old islands. At larger values of eccentricity, the new stable islands keep growing while the old ones shrink. We find that the new stable islands have a size comparable to the old ones when the eccentricity approaches 0.99 (Figure \ref{fig:f2}(d)).

For other values of the mass ratio, $\mu$, the evolution of the surface of section with eccentricity is qualitatively the same, although the sizes of the stable islands vary with $\mu$.  The critical values of the transition eccentricity, $e_c$, also vary but only very modestly with $\mu$.  Table \ref{tab:t1} lists the critical values $e_c$ for the values of $\mu$ that we investigated.

\begin{figure}
\figurenum{3}
\gridline{\fig{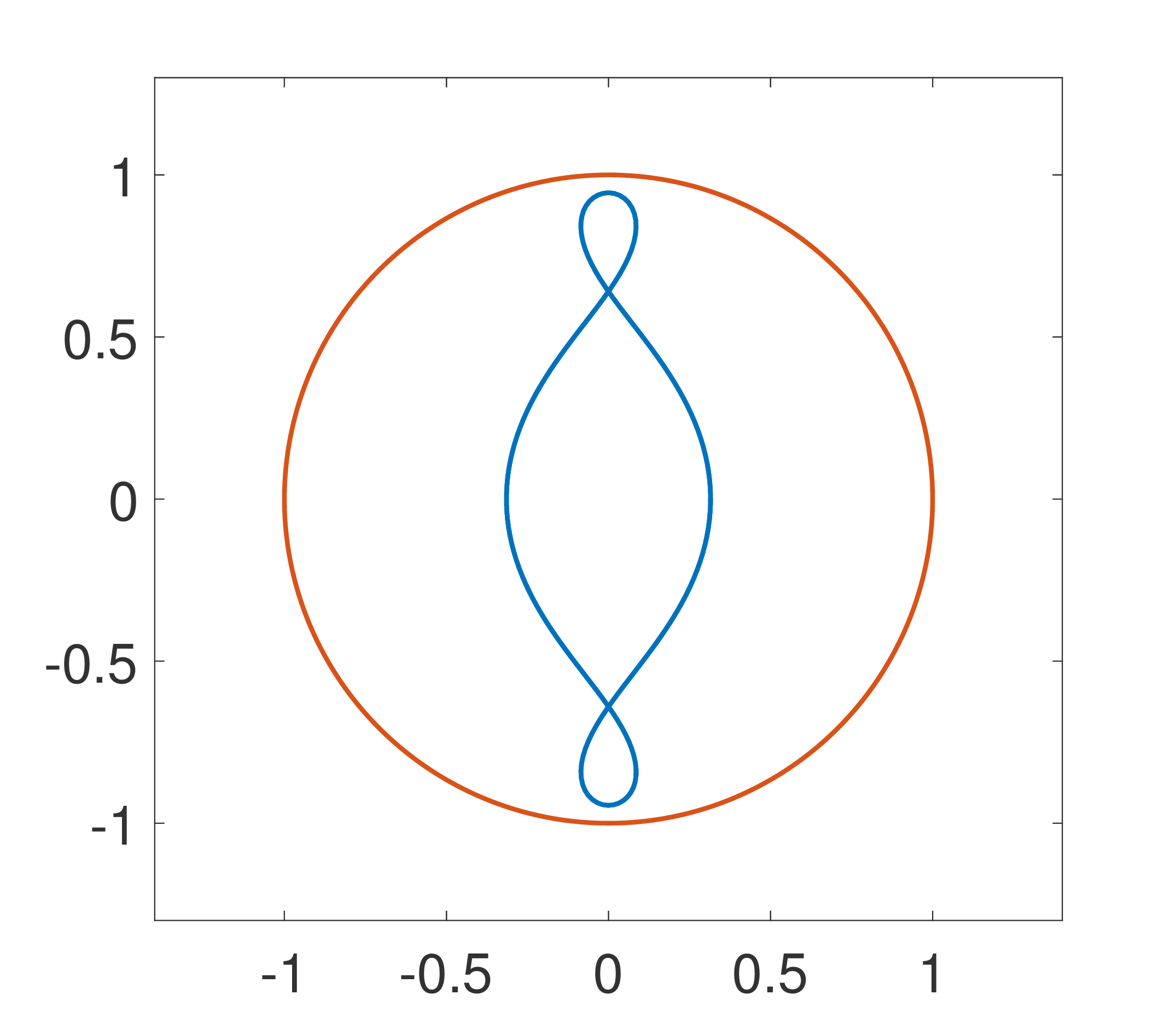}{0.23\textwidth}{(a)}
          \fig{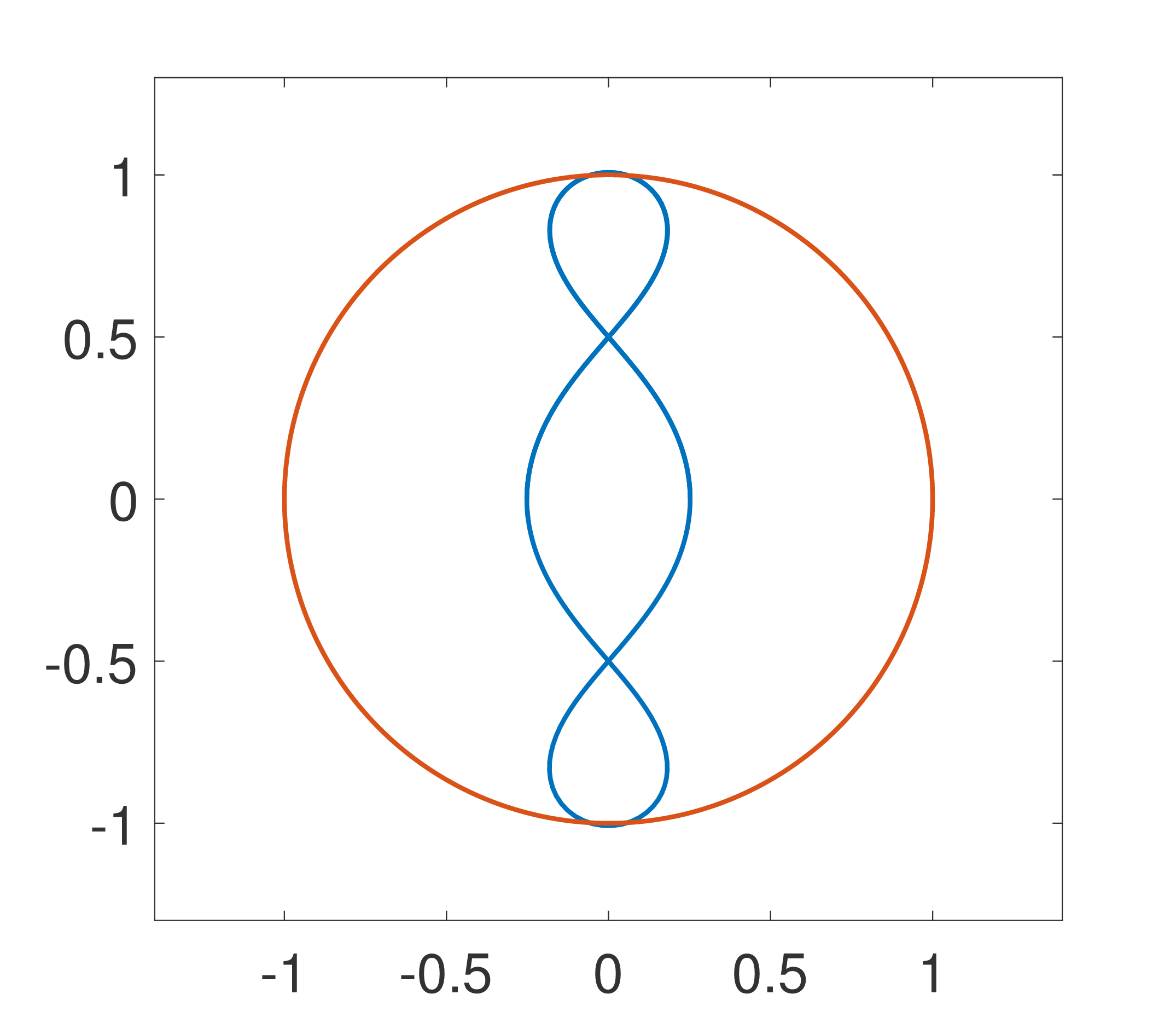}{0.23\textwidth}{(b)}
          \fig{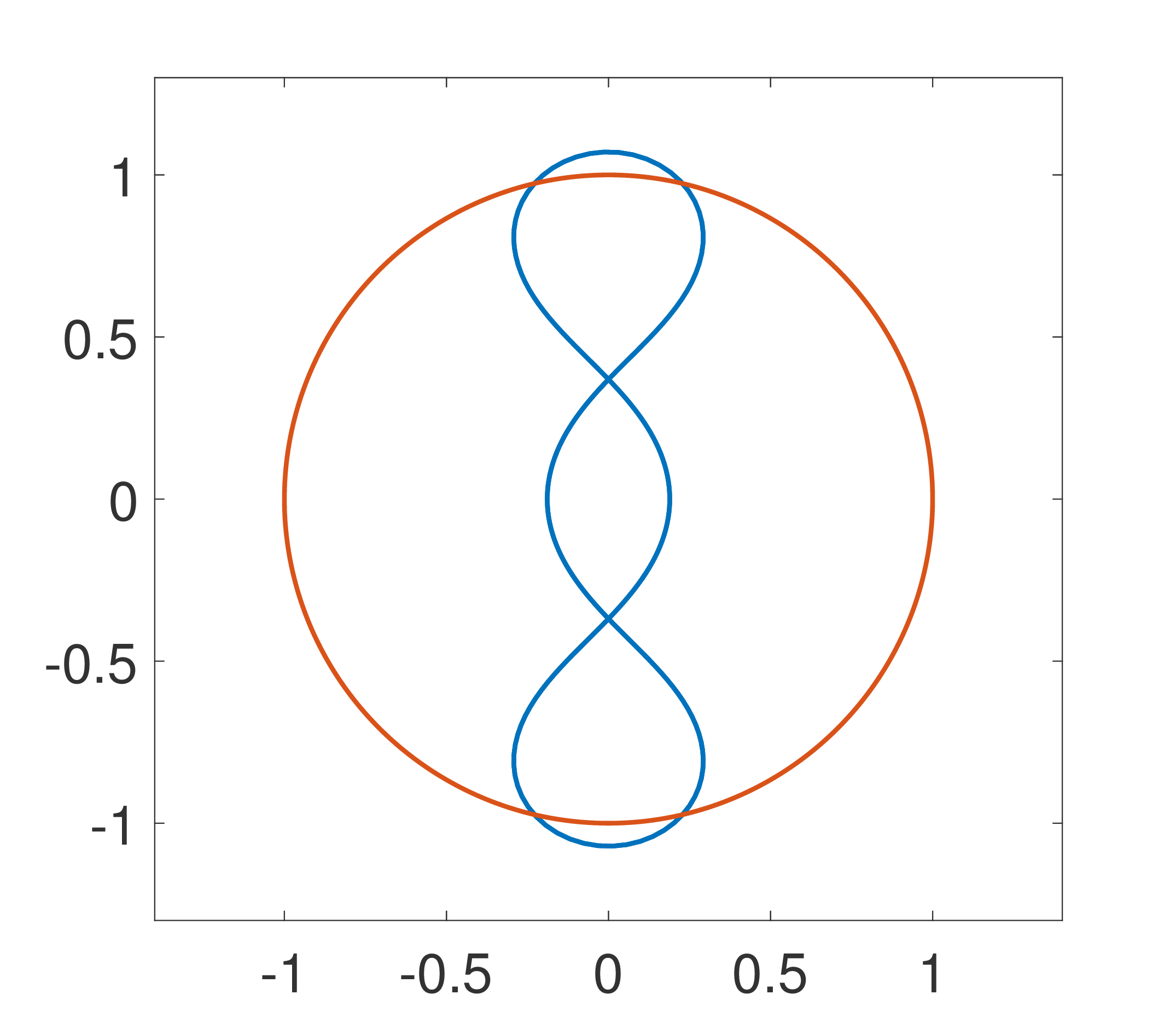}{0.23\textwidth}{(c)}
          \fig{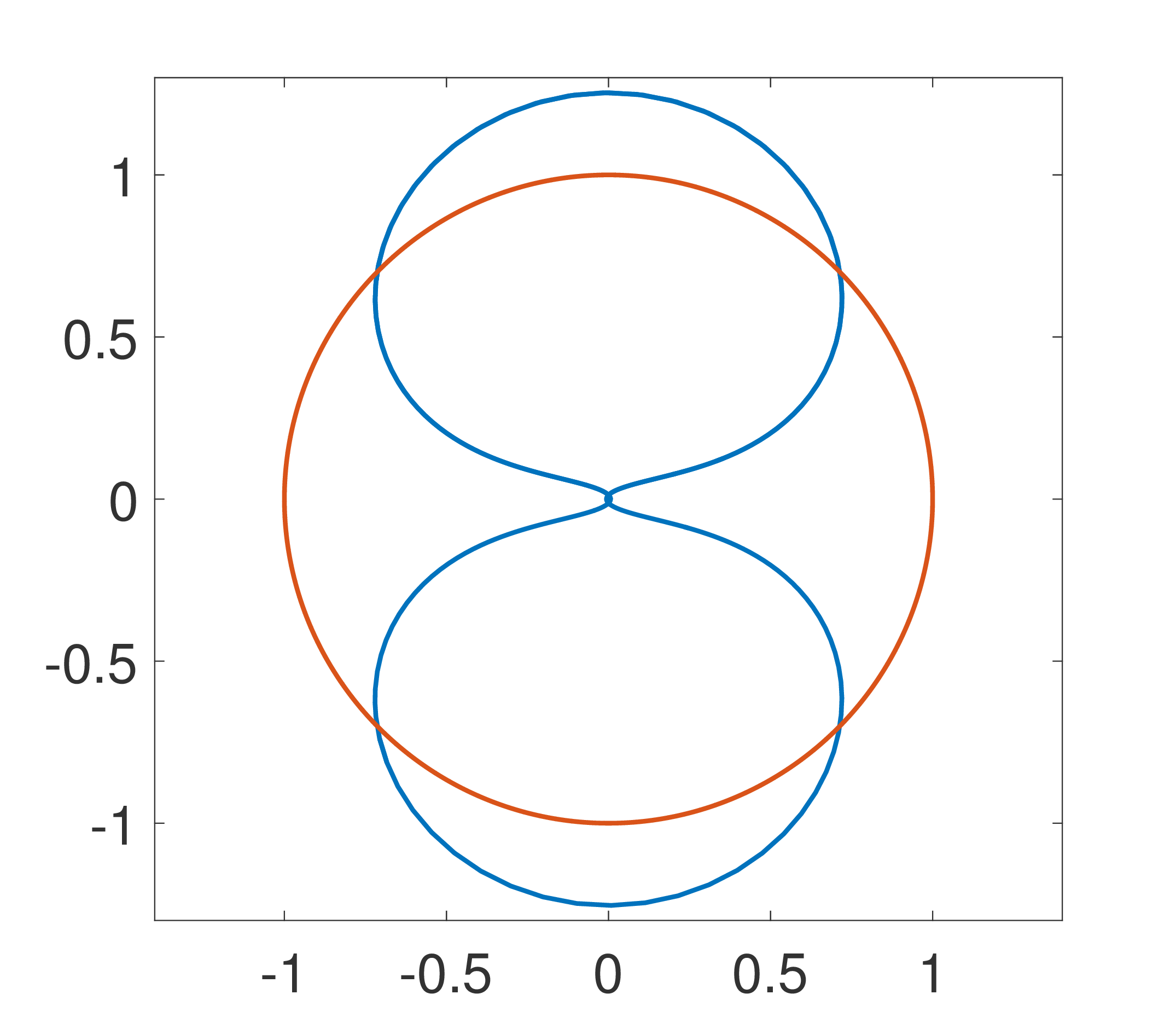}{0.23\textwidth}{(d)}
          }
\caption{Trace of the test particle's 2:1 MMR orbit in the rotating frame. The red circle is the planet's orbit of unit radius. The blue curve is the resonant orbit of the test particle in the rotating frame; there are two pericenter passages represented in this frame, one on the positive abscissa and one on the negative abscissa. From left to right, the test particle's orbit has eccentricity $e=0.50$, $e=0.60$, $e=0.70$ and $e=0.99$, respectively. For eccentricity larger than $0.59$, the test particle's aphelion exceeds the planet's orbit radius.
\label{fig:f3}}
\end{figure}

The transition in the phase space structure with the doubling of the stable islands in the surface of section can be understood by considering the geometry of the orbits in the rotating frame, as illustrated in Figure \ref{fig:f3}.  When the eccentricity is small, the particle's orbit is completely inside the orbit of the planet and its aphelion is far from the orbit of the planet. In this case, the perturbation of the planet is not so large; and there is no chaotic region in the surface of section.  From Figure \ref{fig:f3} we can also see that the test particle's orbit has a two-fold symmetry, so the two centers of the stable islands have a step of $180^\circ$ in $\psi$.  The stable resonance geometry is one in which the planet's angular position is near the abscissa in Figure \ref{fig:f3}, i.e., with $\psi=0$ or $\psi=180^\circ$; in this geometry, the planet's perturbation on the particle is minimized.
At larger eccentricity, the aphelion of the test particle is closer to the orbit of the planet, so the perturbation of the planet on the test particle also becomes larger. The chaos regions appear in the surface of section. At eccentricity larger than $e_{c}=a_{res}^{-1}-1=0.59$, the test particle's orbit is planet-crossing (its aphelion is larger than the planet's orbital radius), and new stable islands appear in the surface of section. Once the aphelion of the test particle's orbit exceeds the orbit radius of the planet, we observe in Figure \ref{fig:f3}(b,c,d) that the two aphelion lobes of the particle's path are cut by the planet's orbit. This is where the new stable islands appear in the surface of section, i.e, centered at $\psi=90^\circ$ and $\psi=270^\circ$. The length of the arc of the planet's orbit which is inside the lobe of the test particle's orbit in the rotating frame reflects the range of $\psi$ for the new stable islands. For larger eccentricity, this length of the arc also becomes larger. So the new stable islands grow and expand at the expense of the old stable islands. For eccentricities near $e=0.99$, the new stable islands are of comparable size with the old ones.

\begin{figure}
\figurenum{4}
\gridline{\fig{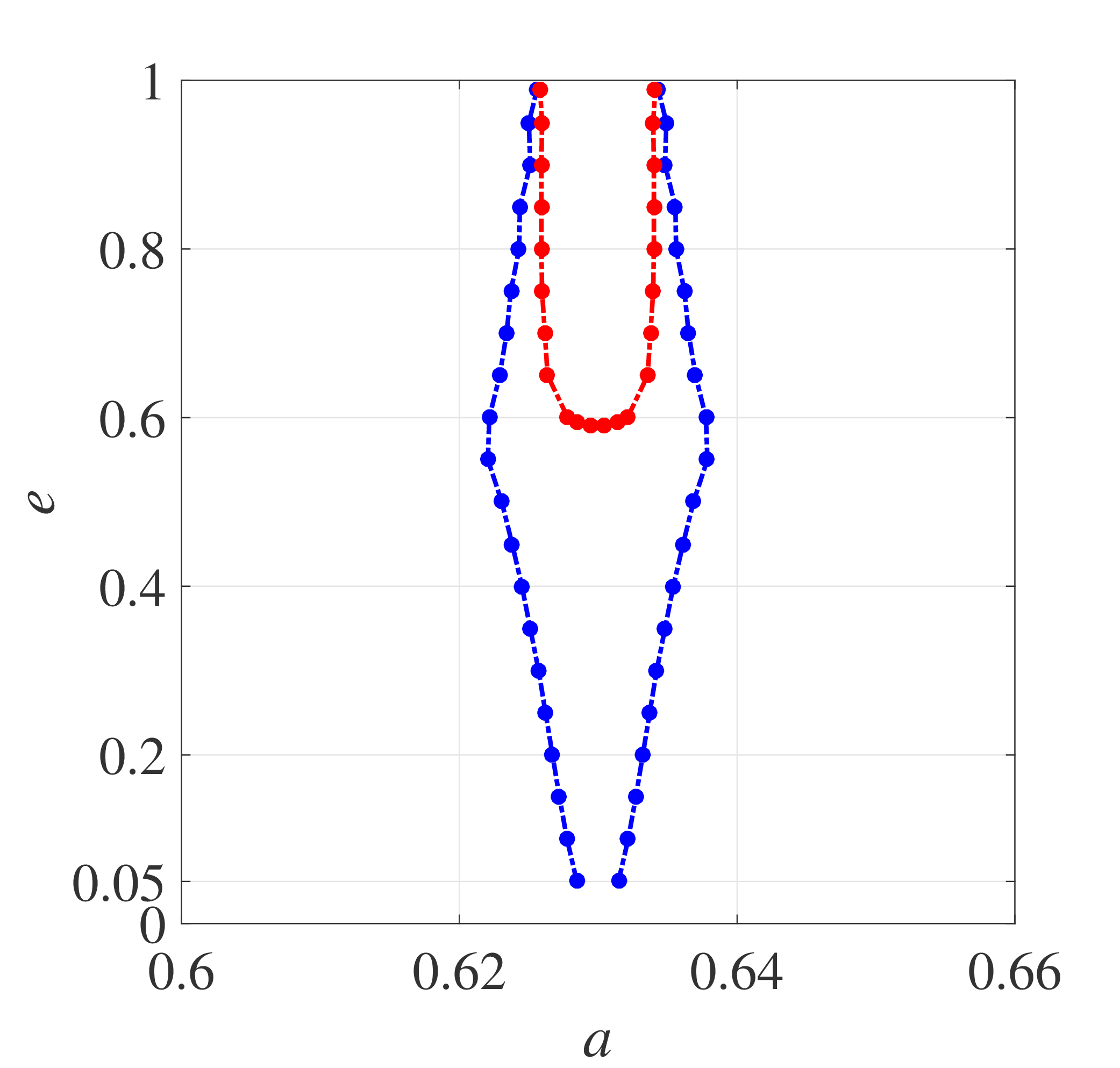}{0.3\textwidth}{(a)}
          \fig{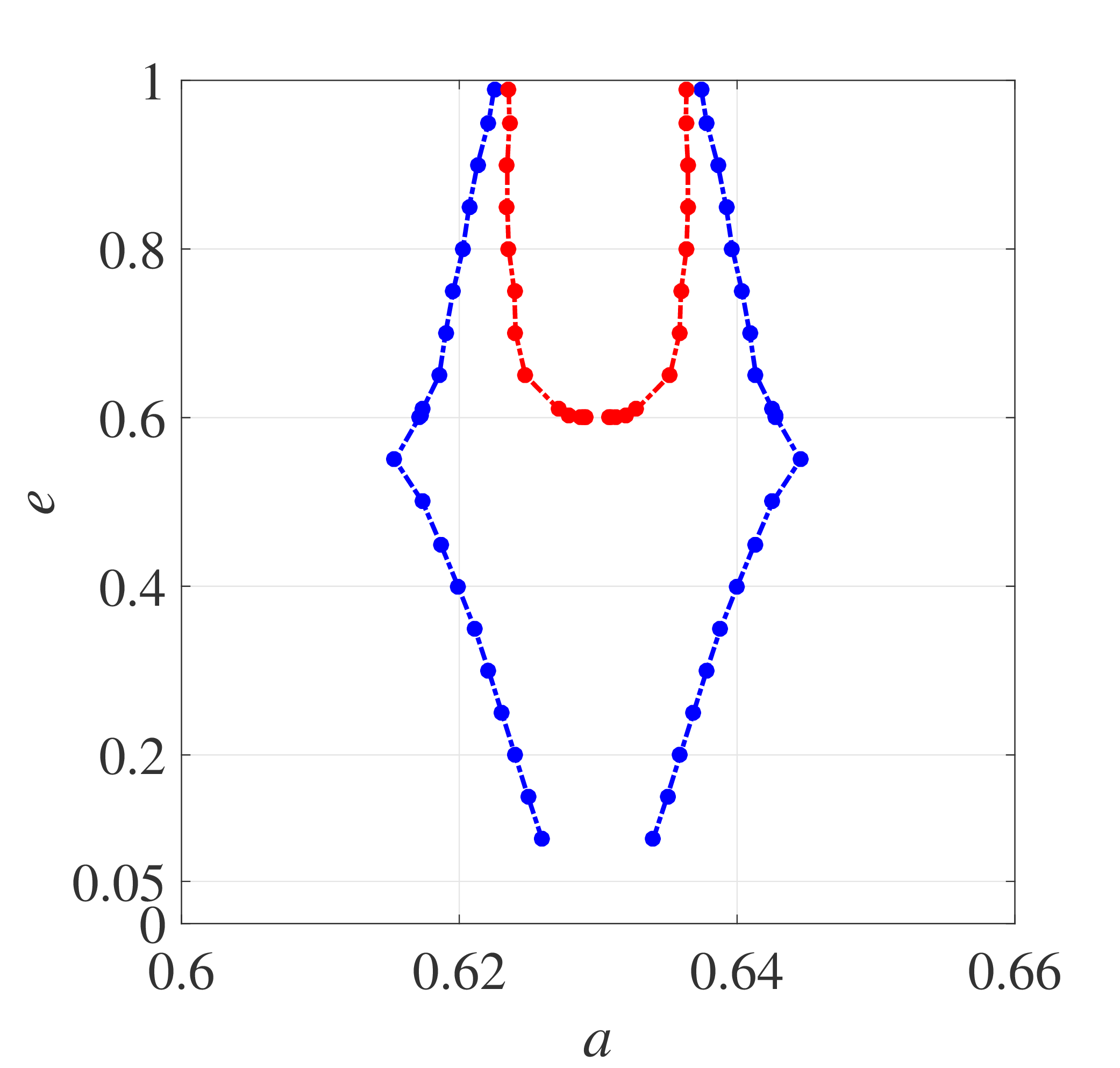}{0.3\textwidth}{(b)}
          \fig{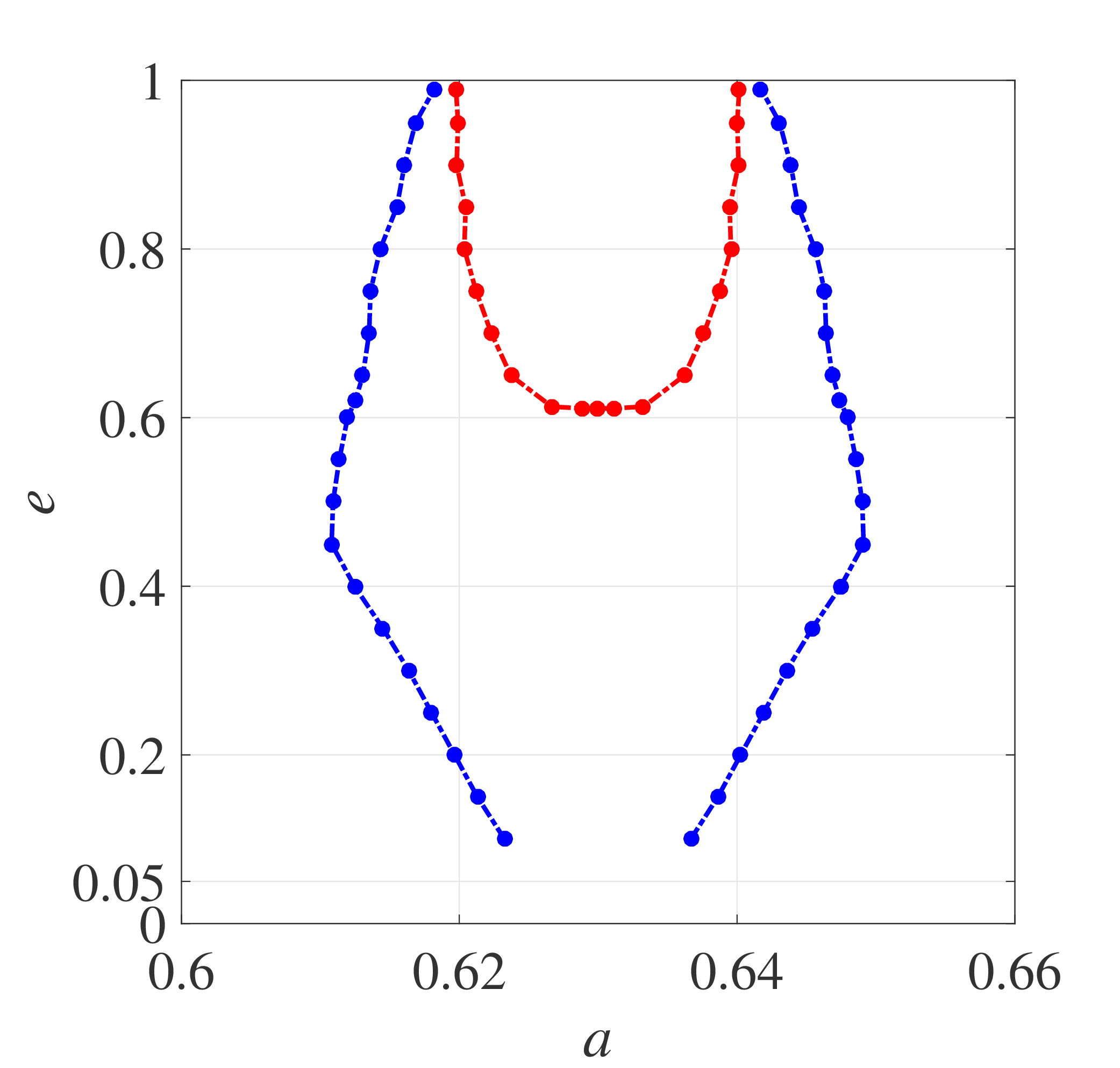}{0.3\textwidth}{(c)}
          }
\caption{Stable resonant libration zone of the 2:1 MMR as a function of eccentricity, for mass ratio $\mu =3 \times 10^{-5}$, $1 \times 10^{-4}$ and $3 \times 10^{-4}$, respectively. The area inside the blue line is the width in semi-major axis for the first resonance zone (with stable islands centered at $\psi=0$ and $180^\circ$). The  area inside the red line is the width in semi-major axis for the second resonance zone (with stable islands centered at $\psi=90^\circ$ and $270^\circ$).
\label{fig:f4}}
\end{figure}

In each surface of section that we generated, we measured the resonance width, $\Delta a$, defined as the maximum range of $a$ about the resonant value, $a_{res}$, for which $\psi$ librates about a stable center.   As noted above, this width depends upon both $\mu$ and the test particle eccentricity.
Figure \ref{fig:f4} plots the width of the 2:1 MMR for test particle eccentricity in the range $0.05$ to $0.99$, for three values of $\mu$. Since there are two dynamically distinct sets of stable islands in the surface of section, we measured them separately. The blue lines plot the widths of the stable islands centered at $\psi=0^\circ, 180^\circ$ in the surface of section; we call these the ``first resonance zone". The red lines plot the widths of the stable islands centered at $\psi=90^\circ, 270^\circ$; these exist for eccentricity exceeding $e_c\simeq0.59$, and we call them the ``second resonance zone".
In these plots, we observe that the first resonance zone increases in width with increasing eccentricity up to a maximum width, and then decreases with increasing eccentricity. The maximum width of the first resonance zone is an increasing function of the mass ratio, $\mu$. Table \ref{tab:t1} lists the values of the eccentricity, $e_m$, where the first resonance zone has its maximum width, for each mass ratio $\mu$ that we investigated. We find that $e_m$ decreases with increasing $\mu$, i.e., the maximum width occurs at smaller eccentricity for larger $\mu$.  In Figure \ref{fig:f4} we can see that the first resonance zone reaches its maximum width just before the second resonance zone appears.  The width of the second resonance zone, once it appears, increases rapidly initially but then more gradually with increasing eccentricity; it reaches maximum width as the eccentricity approaches $e=0.99$. We can also note that the condition for the existence of the second resonance zone for $e>e_c\simeq0.59$, depends mainly on the orbital geometry, and only weakly on the mass ratio $\mu$.

\floattable
\begin{deluxetable}{ccc}
\tablecaption{Values of $e_m$ and $e_c$ for the interior 2:1 MMR  \label{tab:t1}}
\tablehead{
\colhead{Mass ratio, $\mu$} & \colhead{$e_m$}  & \colhead{$e_c$}
}
\startdata
  $1 \times 10^{-5}$ & $0.59$ & 0.590\\
  $3 \times 10^{-5}$     & $0.58$  &0.593 \\
  $1 \times 10^{-4}$     & $0.57$   &0.602 \\
  $3 \times 10^{-4}$     & $0.48$    &0.612\\
  $1 \times 10^{-3}$     & $0.42$  &0.648 \\
  $3 \times 10^{-3}$     & $0.31$   &0.709\\
\enddata
\tablecomments{$e_m$ represents the eccentricity where the first resonance zone has its maximum width, $\Delta a$; $e_c$ is the value of the eccentricity above which the second resonance zone is present in the phase space.
}
\end{deluxetable}

\begin{deluxetable}{ccccc}
\tablecaption{Best-fit power-law parameters of the width of the first and second resonance zones of the interior 2:1 MMR \label{tab:t2}}
\tablehead{
\colhead{} & \colhead{First zone} & \colhead{} & \colhead{Second zone} & \colhead{} \\
\colhead{Eccentricity} & \colhead{$\alpha$} & \colhead{$\beta$} & \colhead{$\alpha$} & \colhead{$\beta$}
}
\startdata
0.1 & 0.7607          & 0.4950           & --          & -- \\
0.2 & 1.183           & 0.4994  & --          & -- \\
0.3 & 1.621           & 0.5035  & --          & -- \\
0.4 & 2.066           & 0.5031           & --          & -- \\
0.5 & 0.9990          & 0.4051           & --          & -- \\
0.6 & 0.9766          & 0.3969  & --          & -- \\
0.7 & 0.8475          & 0.4009           & 0.2248  & 0.3216 \\
0.8 & 0.7087          & 0.3929           & 0.2418  & 0.3233  \\
0.9 & 0.7268          & 0.4063  & 0.3728  & 0.3667  \\
\enddata
\tablecomments{The second resonance zone does not exist for $e \lesssim 0.6$.
}
\end{deluxetable}

To simplify the quantification of the mass dependence of the resonance widths, $\Delta a$, of the first and second resonance zones, we postulate that, in each case, $\Delta a$ has a power law relation with the mass of the planet, and we empirically fit the values of $\Delta a$ (measured from the surfaces of section) to such a function,
\begin{equation}\label{eq22}
\Delta a=\alpha\mu^{\beta},
\end{equation}
allowing that the parameters, $\alpha$ and $\beta$, depend upon the test particle eccentricity.

\begin{figure}
\figurenum{5}
\gridline{\fig{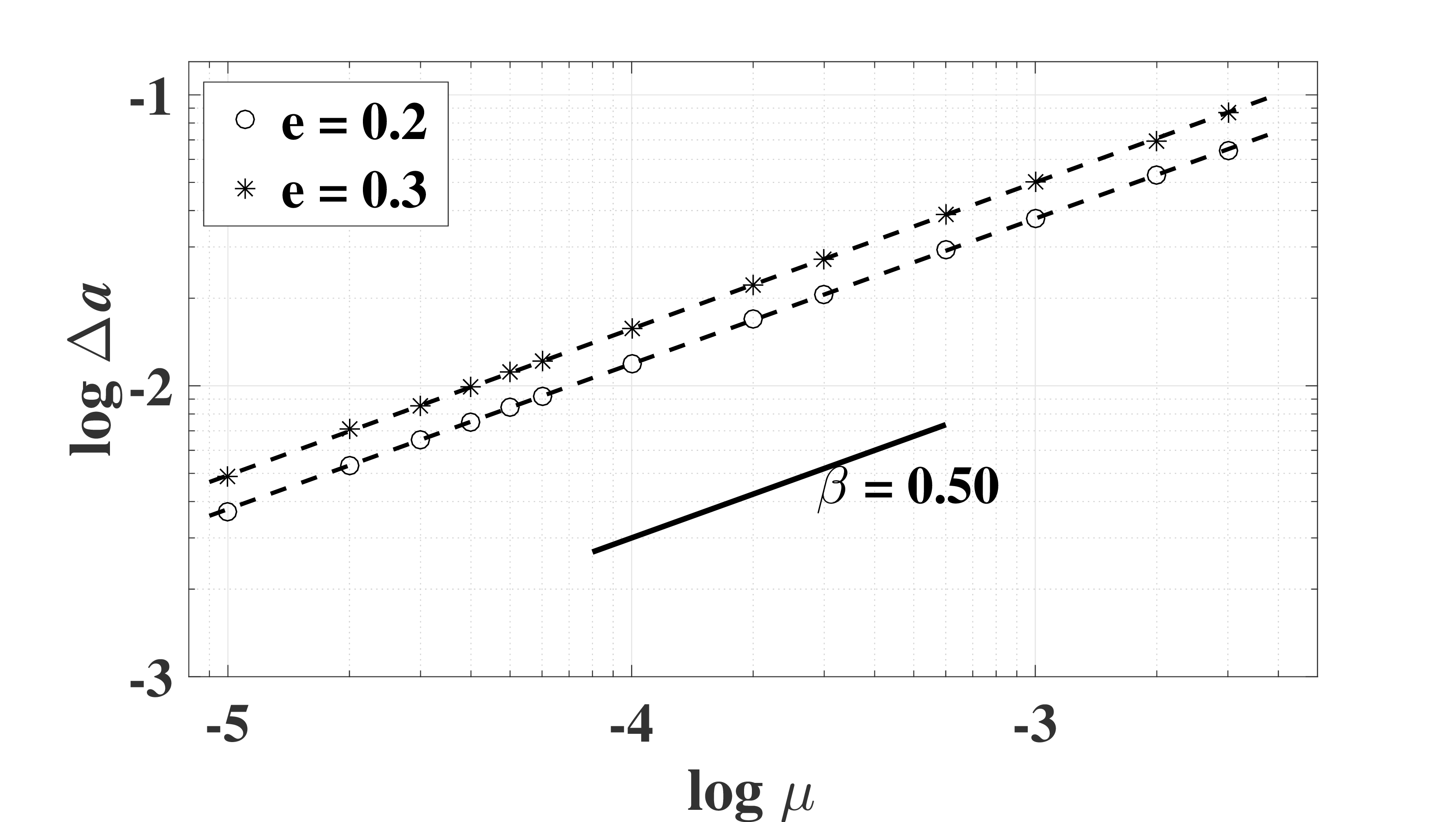}{0.5\textwidth}{(a)}
          \fig{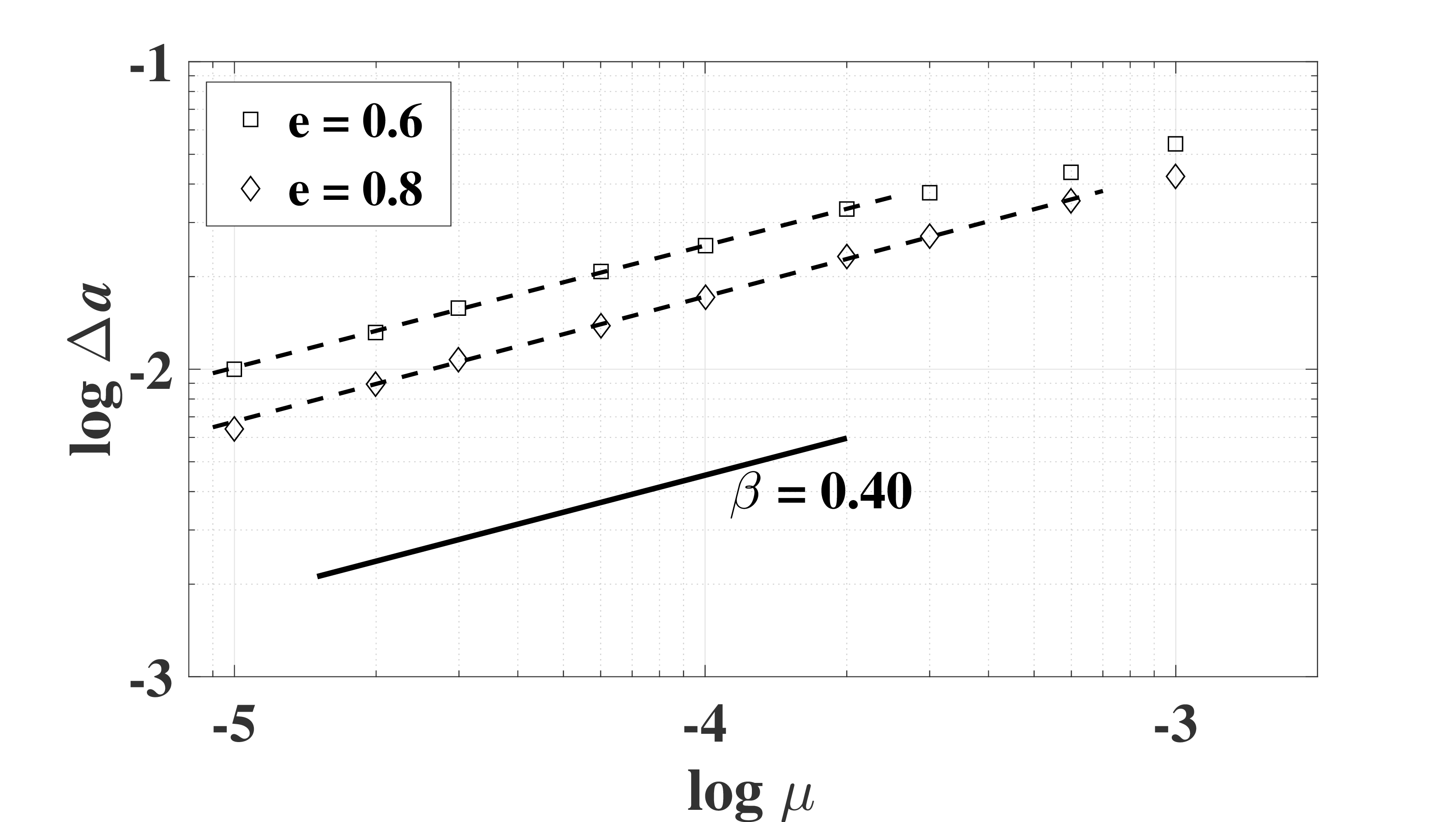}{0.5\textwidth}{(b)}
          }
\caption{Planet-mass dependence of the width, $\Delta a$, of the 2:1 MMR's ``first resonance zone": (a) for low eccentricities, $e=0.2$ (circles), $e=0.3$ (stars), and (b) for high eccentricities, $e=0.6$ (squares), $e=0.8$ (diamonds). The dashed lines are the best-fit power-law relation between $\Delta a$ and mass ratio $\mu$.  The black solid line is a reference line with the indicated slope value.
\label{fig:f5}}
\end{figure}

\begin{figure}
\figurenum{6}
\plotone{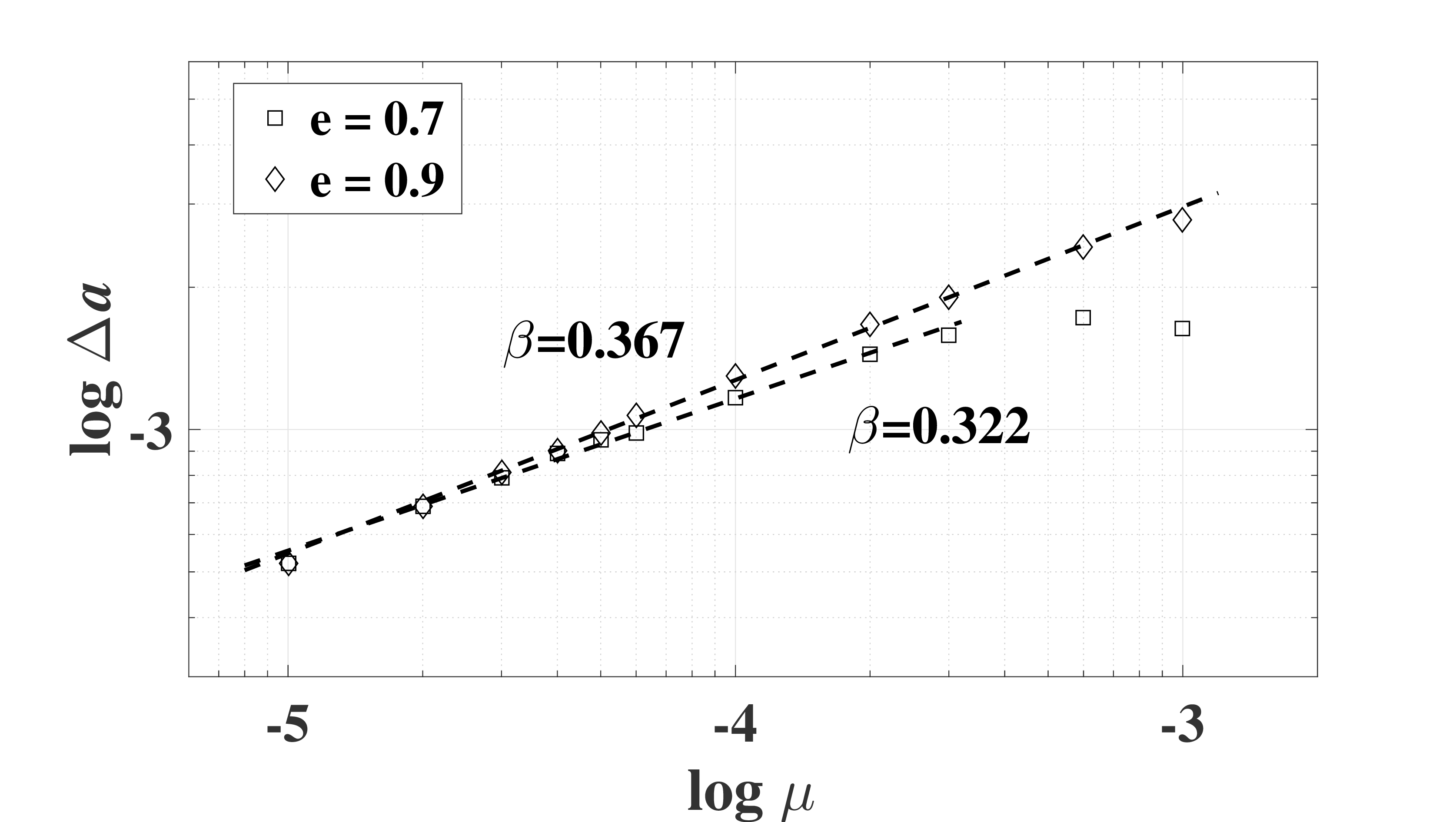}
\caption{Planet-mass dependence of the width, $\Delta a$, of the 2:1 MMR's ``second resonance zone", for eccentricity $e=0.7$ (squares) and $e=0.9$ (diamonds). The dashed lines are the best-fit power-law relation between $\Delta a$ and mass ratio $\mu$.
\label{fig:f6}}
\end{figure}

In Figure \ref{fig:f5} we plot the stable libration width, $\Delta a$, of the first resonance zone as a function the mass ratio $\mu$, for several fixed values of the eccentricity.  For the lower eccentricities, $e=0.2$ and $e=0.3$, the aphelion of the test particle does not exceed the orbit radius of the planet and only the first resonance zone exists.
In this low eccentricity regime, for small mass ratio, the width of stable resonance zones increases with increasing eccentricity; the boundaries of the stable islands in the surfaces of section are very smooth and no chaotic region exists; however, for large mass ratio, $\mu\gtrsim(2-3)\times10^{-4}$, chaotic regions appear in the surfaces of section but the power law relation still holds.
The best-fit power-law relations are found to be $\Delta a=1.183\mu^{0.4994}$ and $\Delta a=1.621\mu^{0.5035}$ for $e=0.2$ and $e=0.3$, respectively; the power-law index is near $1/2$.
For the larger eccentricities, $e=0.6$ and $e=0.8$, the aphelion of the particle exceeds the orbit radius of the planet, and the stable islands in the surfaces of section are surrounded by chaotic regions even for small mass ratio. In this large eccentricity regime, the first resonance zone width shrinks with increasing eccentricity.
The best-fit power-law relations are $\Delta a=0.9766\mu^{0.3969}$ and $\Delta a=0.7087\mu^{0.3929}$ for $e=0.6$ and $e=0.8$, respectively; the power-law index is near 2/5.

In Figure \ref{fig:f6} we plot the stable libration width, $\Delta a$, of the second resonance zone as a function the mass ratio $\mu$, for $e=0.7$ and $e=0.9$. In this eccentricity regime, the second resonance zone's width is not strongly varying with eccentricity (as can be seen in  Figure \ref{fig:f4}). The best-fit power-law relations are found to be $\Delta a=0.2248\mu^{0.3216}$ and $\Delta a=0.3728\mu^{0.3667}$ for $e=0.7$ and $e=0.9$, respectively.
The power-law index is larger for the higher eccentricity.
We note that for eccentricity near $0.6$--$0.7$, there is a rollover in the slope at larger $\mu$, $\mu\gtrsim3\times10^{-4}$ (Figure~\ref{fig:f5}(b) and Figure~\ref{fig:f6}).  This rollover is associated with the existence of significant chaotic regions surrounding the resonance zones.

Table \ref{tab:t2} summarizes the results for the best-fit power-law relations for the widths of the first and second resonance zones for different eccentricities.  The dependence on the eccentricity is not described well as a simple power law or even a piece-wise power law, and we have not attempted to fit a functional form for it.

\subsection{Interior 3:2 MMR\label{subsec:3to2}}

A test particle near the 3:2 MMR with the planet has a semi-major axis
\begin{equation}\label{eq23}
a_{res}=\left(\frac{2}{3}\right)^{2/3}=0.763.
\end{equation}
For an elliptic orbit, the maximum of the aphelion is
\begin{equation}\label{eq24}
\lim_{e \to 1} a_{res}\left(1+e\right)=1.526,
\end{equation}
therefore, the test particle's orbit is planet-crossing when the eccentricity exceeds $a_{res}^{-1}-1\simeq0.31$.

\begin{figure}
\figurenum{7}
\gridline{\fig{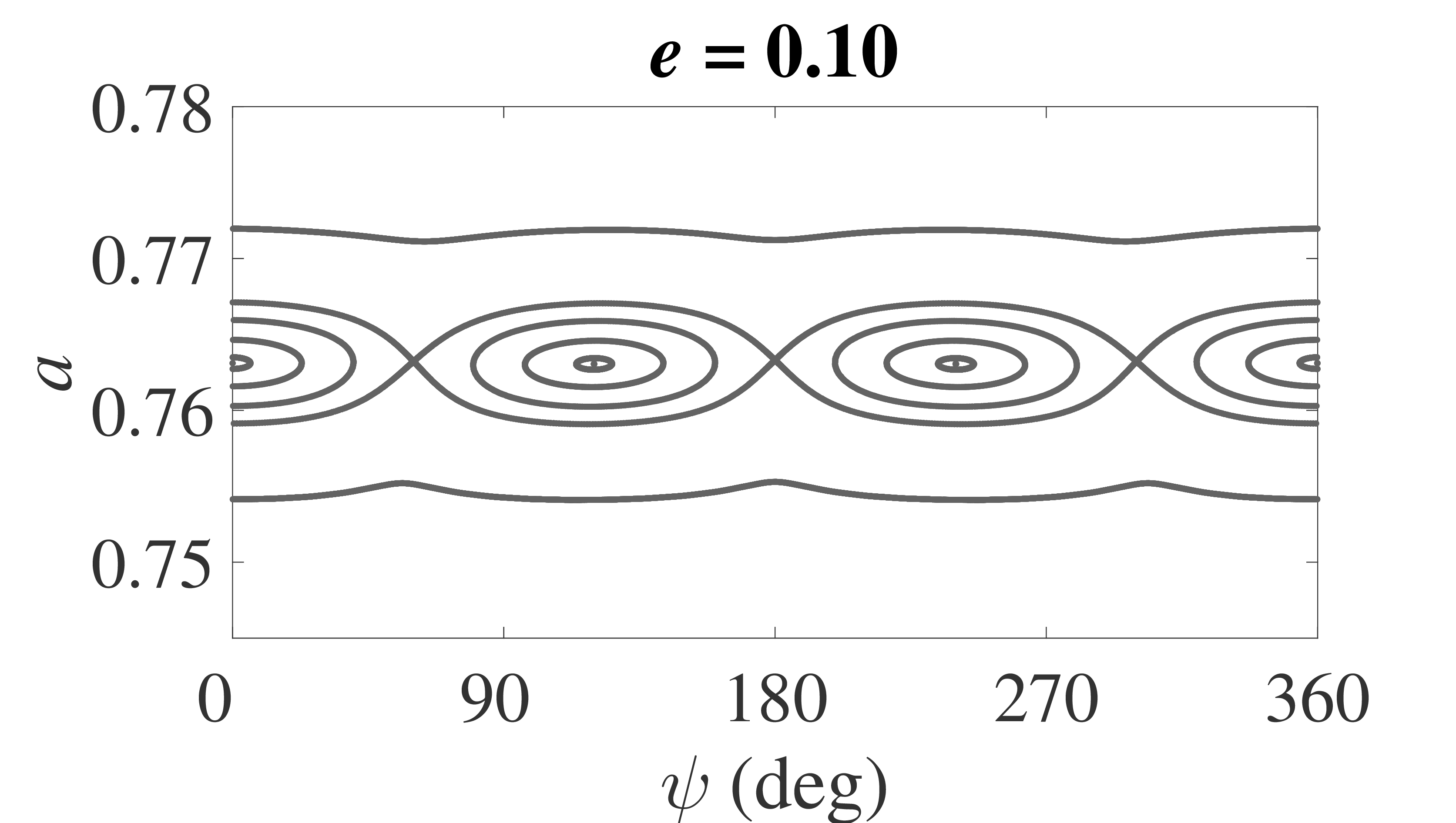}{0.45\textwidth}{(a)}
          \fig{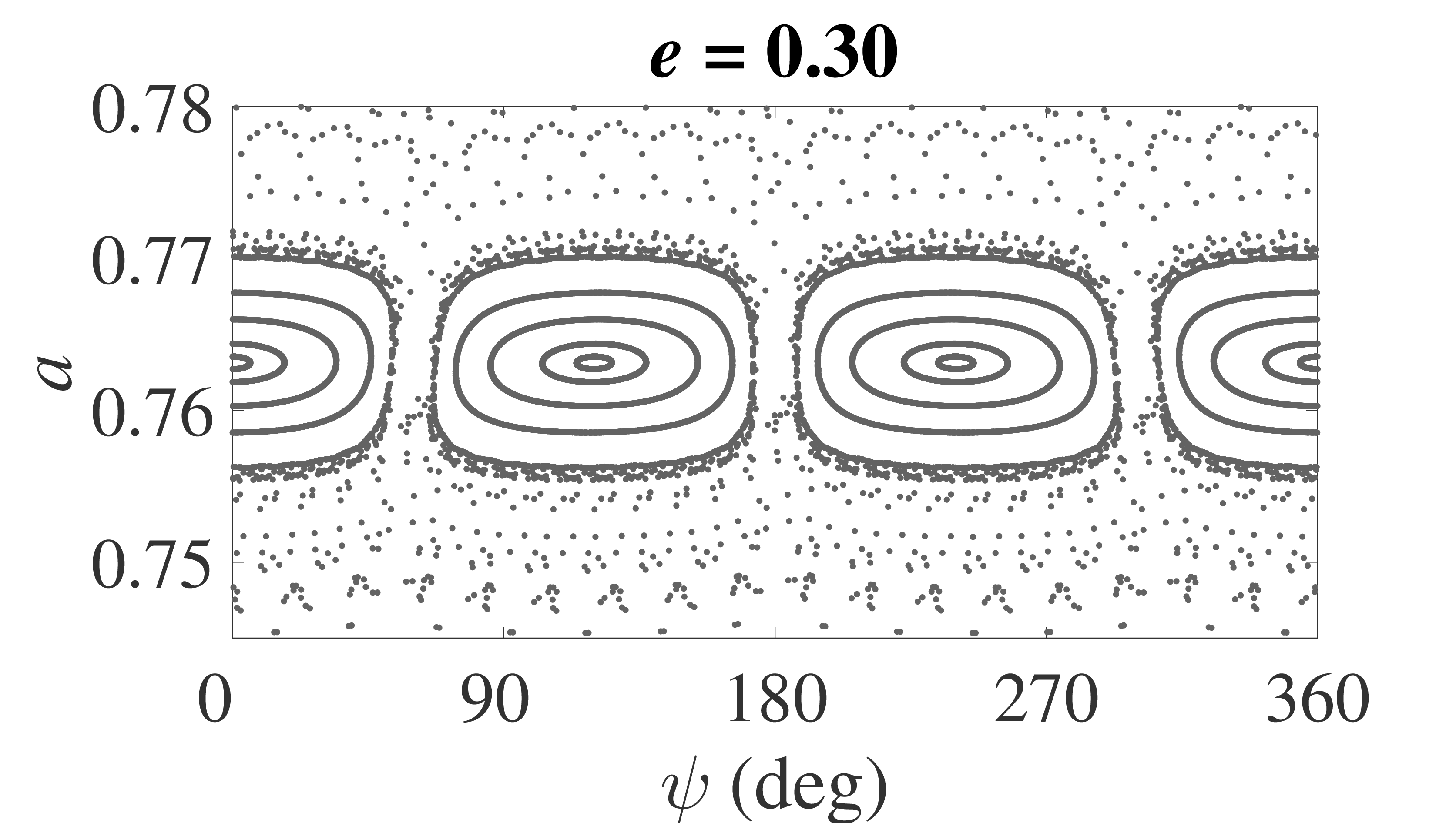}{0.45\textwidth}{(b)}
          }
\gridline{\fig{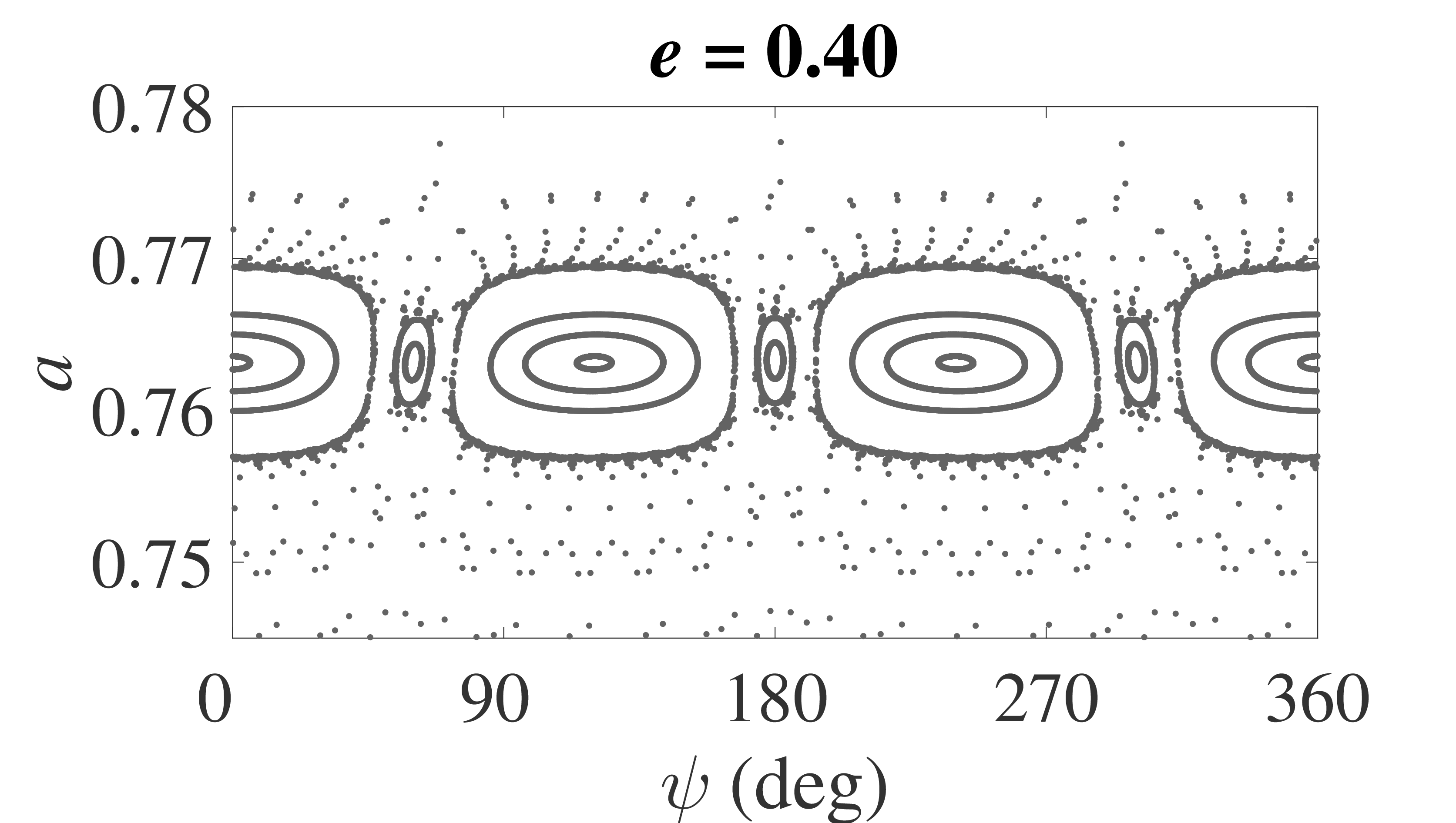}{0.45\textwidth}{(c)}
          \fig{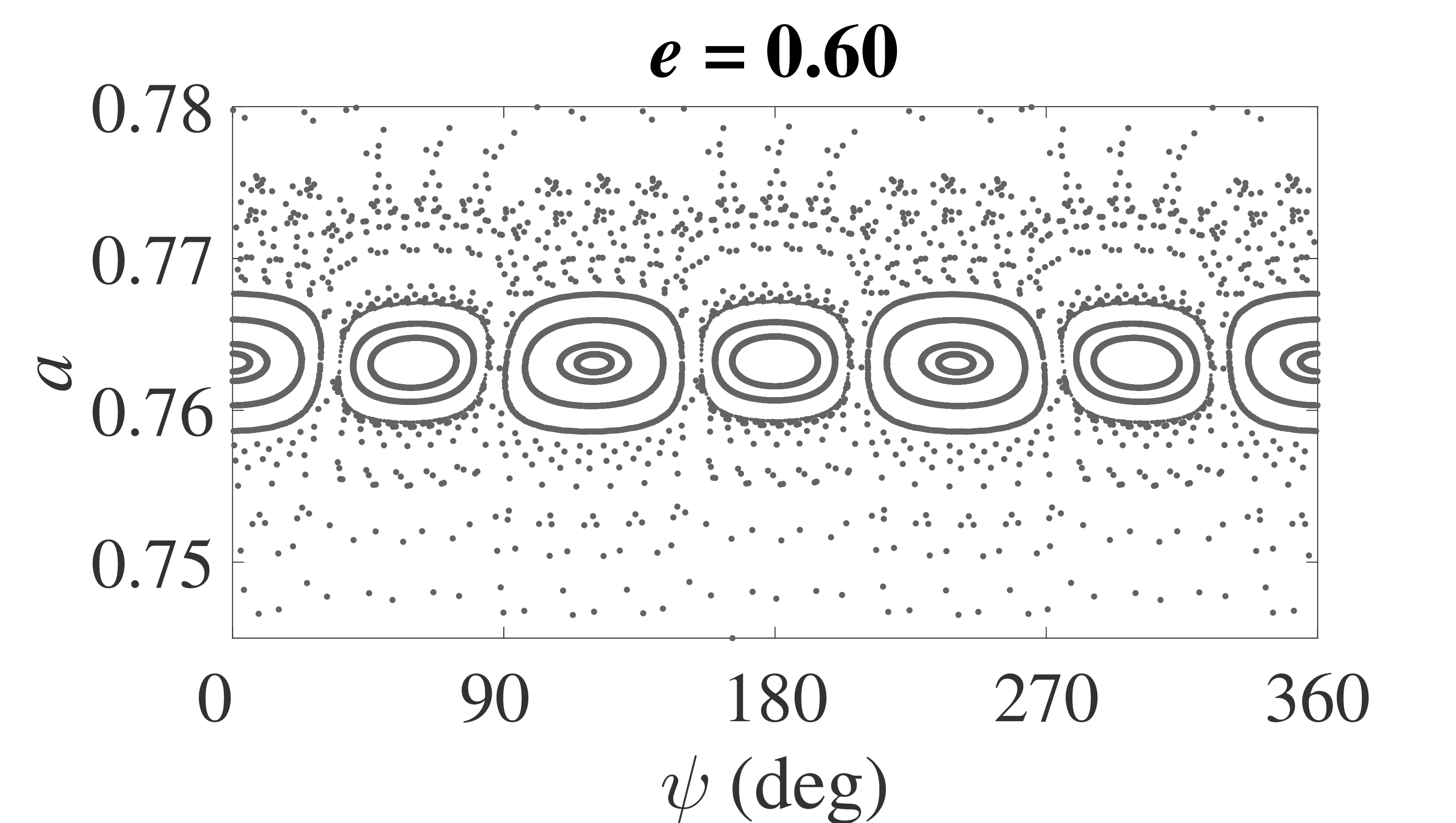}{0.45\textwidth}{(d)}
          }
\gridline{\fig{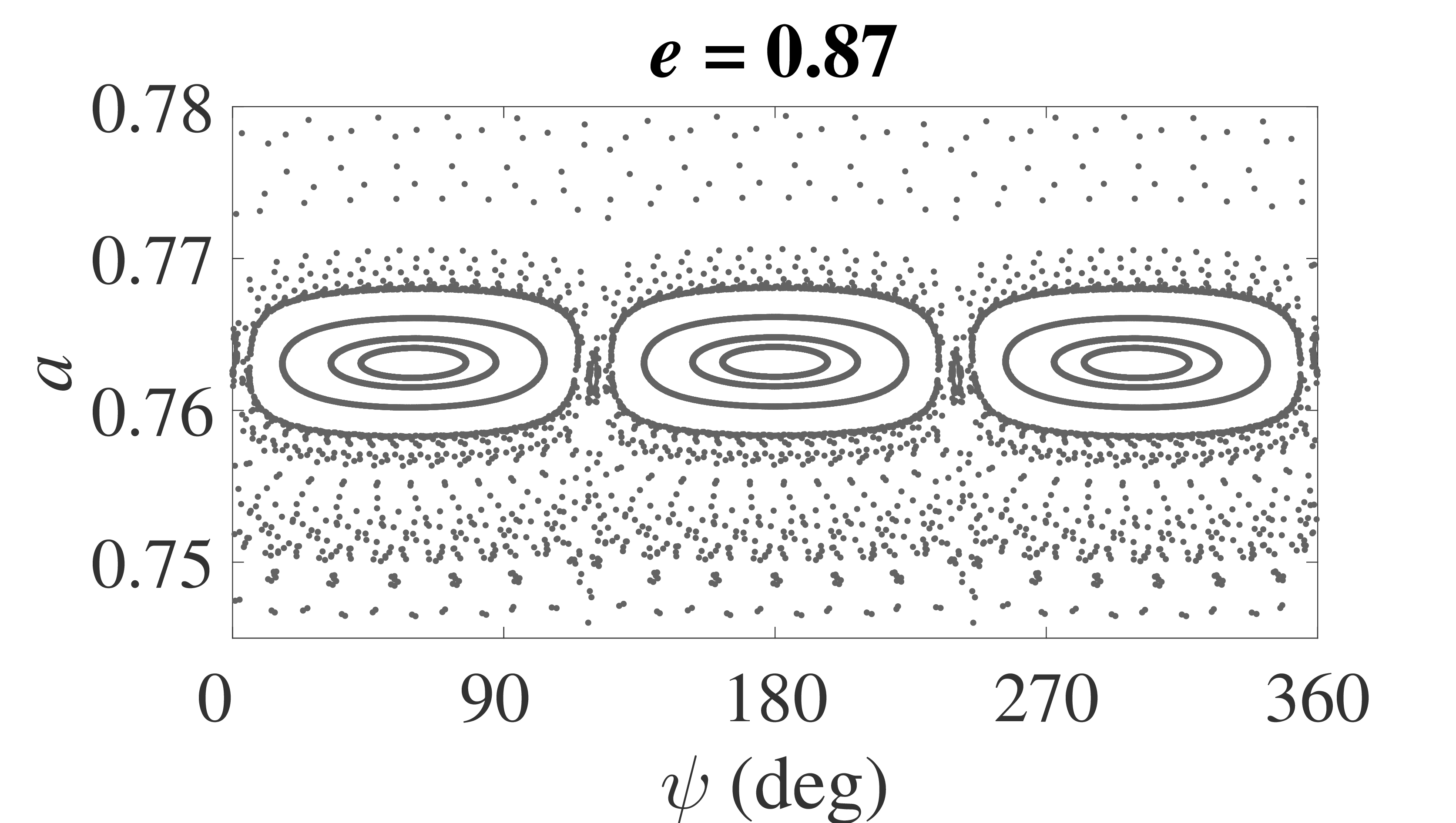}{0.45\textwidth}{(e)}
          \fig{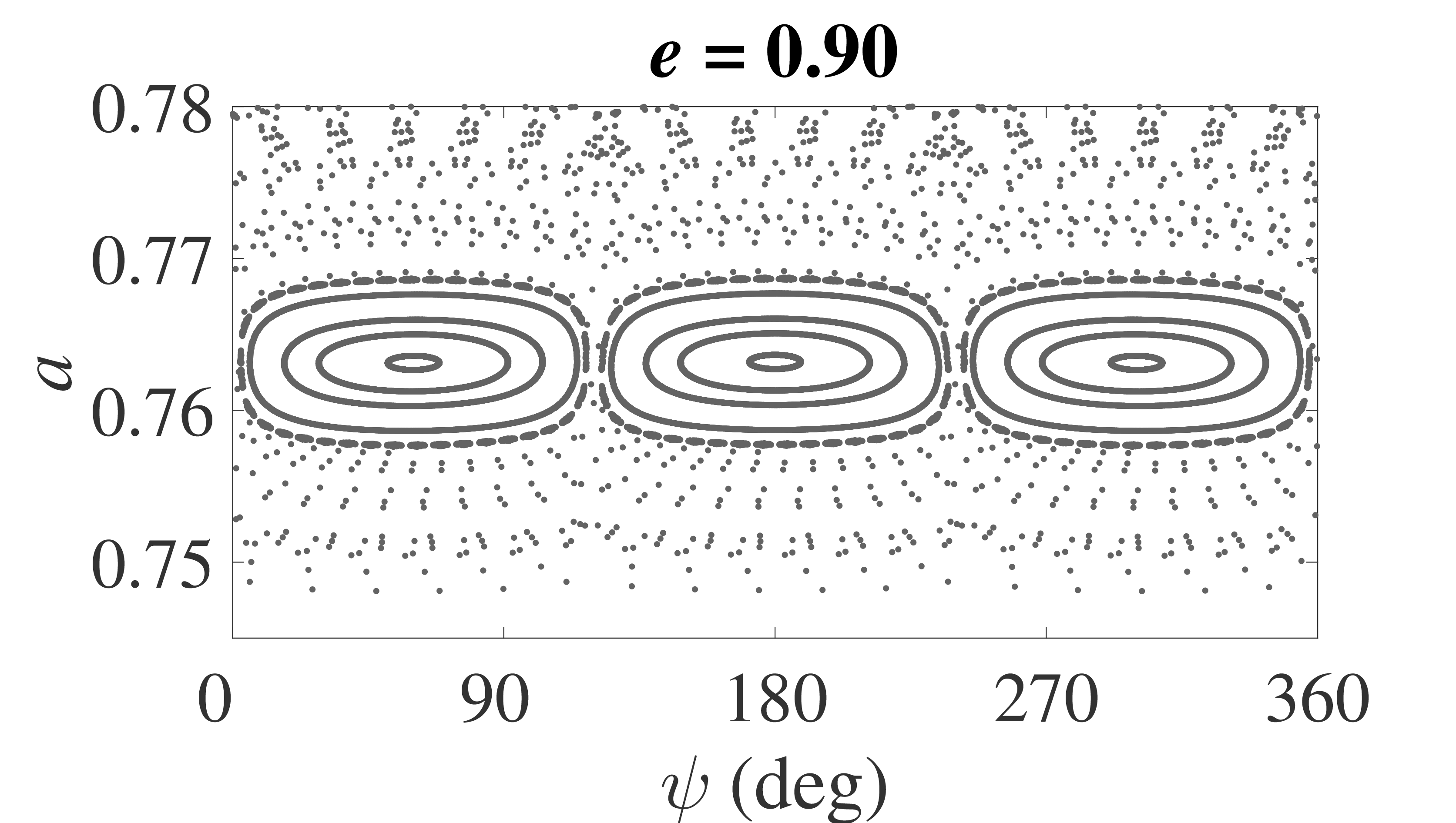}{0.45\textwidth}{(f)}
          }
\gridline{\fig{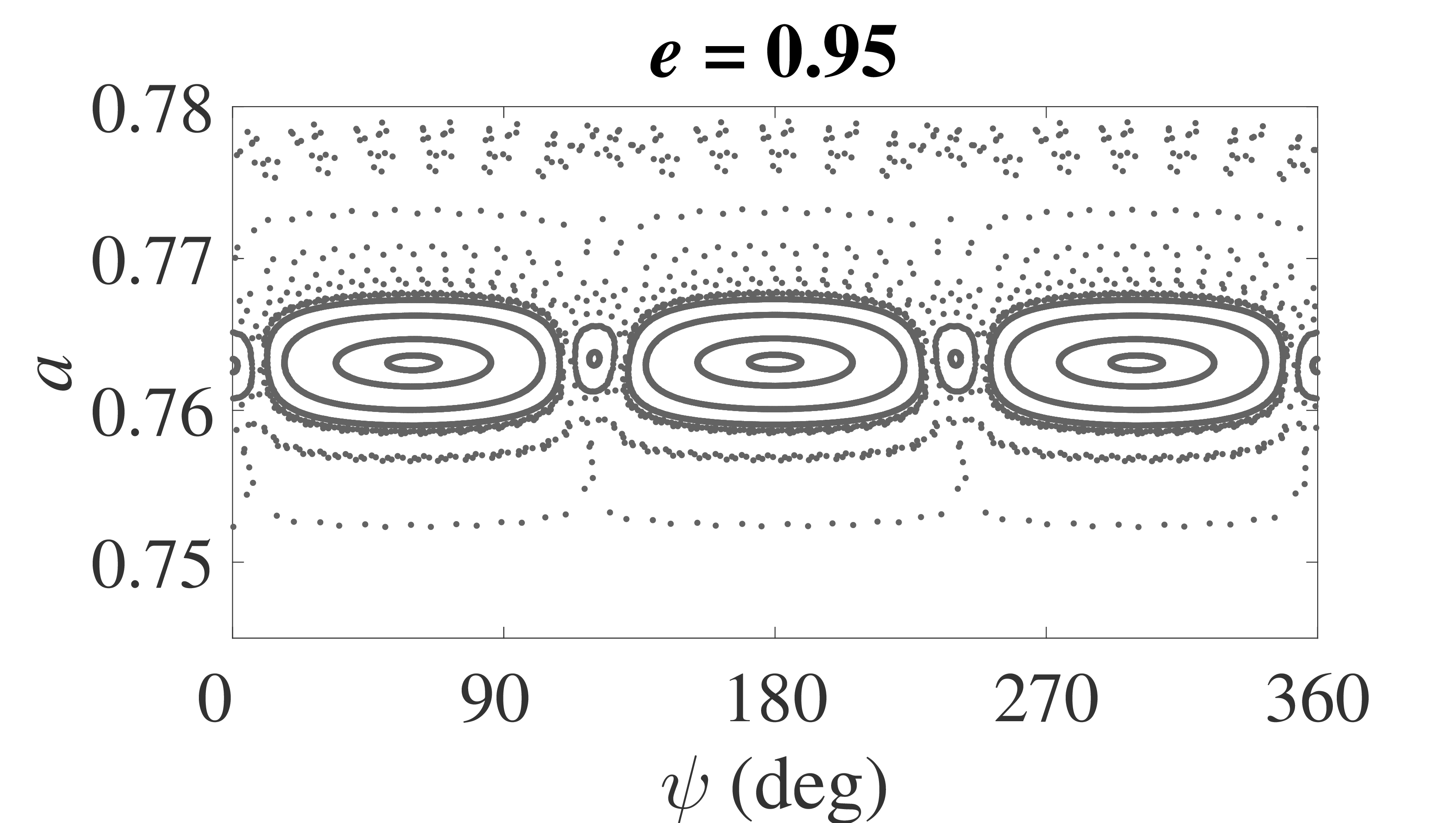}{0.45\textwidth}{(g)}
          \fig{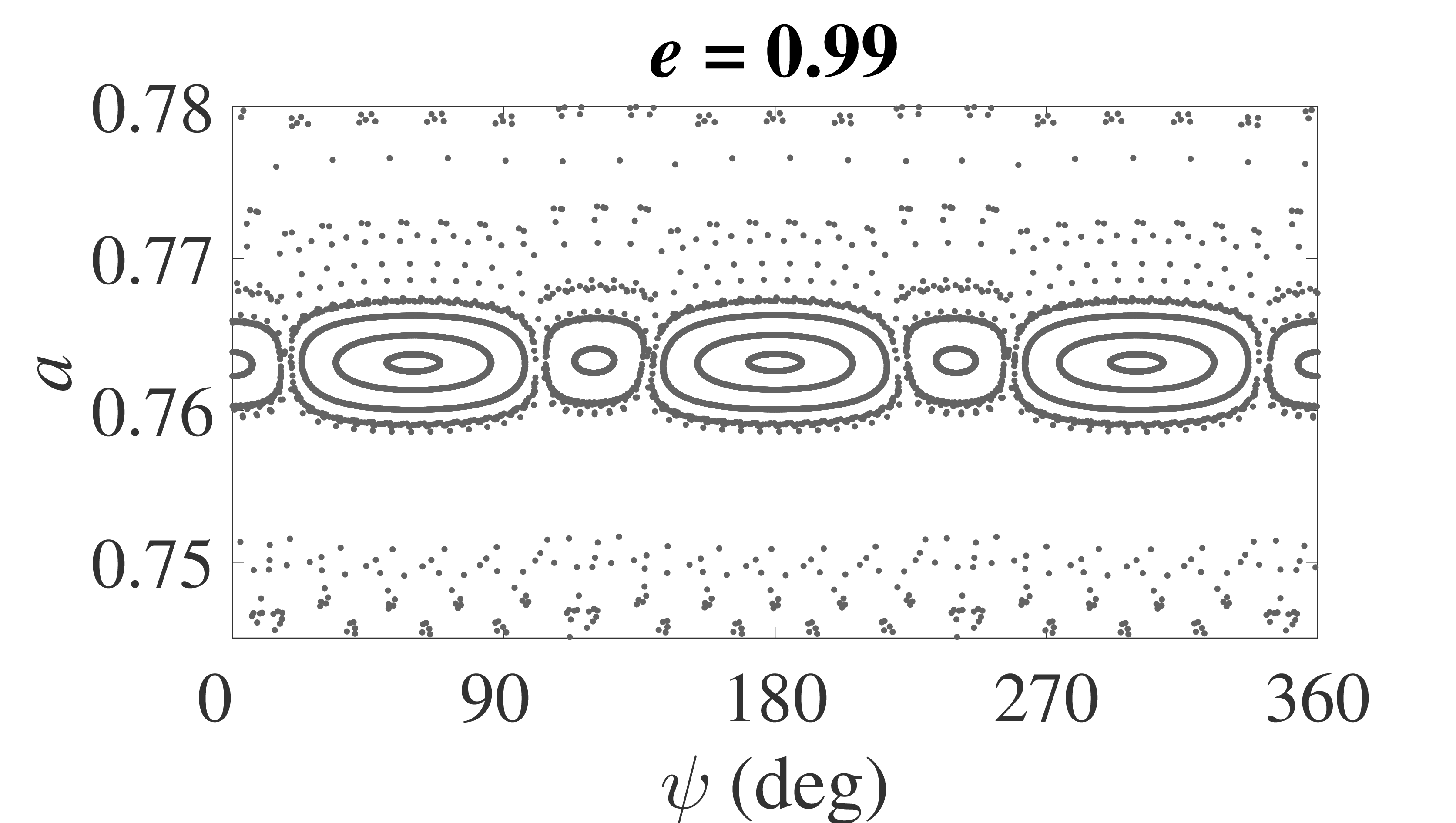}{0.45\textwidth}{(h)}
          }
\caption{Surfaces of section in $(a,\psi)$ near the interior 3:2 MMR , for mass ratio $\mu = 3 \times 10^{-5}$.
Each panel has a fixed value of the Jacobi constant, parametrized by the combination of the resonance value of the semi major axis, $a_{res}=0.763$, and the indicated value of the eccentricity.  \label{fig:f7}}
\end{figure}

We computed the surfaces of section for the 3:2 MMR for $0.05\leq e \leq 0.99$ and for $1\times10^{-5}\leq \mu \leq 3\times10^{-3}$.  Results are shown in Figure \ref{fig:f7} for $\mu =3 \times 10^{-5}$. We can see that the 3:2 MMR phase space is more complicated than the 2:1 case. There are three transitions in the evolution of the surfaces of section with increasing eccentricity. When the eccentricity is small (Figure \ref{fig:f7}(a)), there are three stable islands in the surface of section, which are centered at $\psi=0^\circ$, $120^\circ$ and $240^\circ$; their boundaries are smooth and no chaotic regions exist. As the eccentricity approaches the first critical value, $e_{c1}=0.33$, the widths of these three stable islands grow and chaotic regions appear and expand in the surface of section (Figure \ref{fig:f7}(b)). When the eccentricity exceeds $e_{c1}$, the number of stable islands doubles from three to six (Figure \ref{fig:f7}(c)); the three new stable islands appear in the surface of section centered at $\psi=60^\circ$, $180^\circ$ and $300^\circ$.
At larger values of eccentricity, the new stable islands keep growing while the old ones shrink. From Figure \ref{fig:f7}(d) we can see that when the eccentricity is 0.60, the new stable islands have a size comparable with the old ones. As the eccentricity increases further, the old stable islands keep shrinking and the new ones keep growing. When the eccentricity exceeds a second critical value, $e_{c2}$, the old stable islands disappear and only the new islands exist in the surface of section. The number of the stable islands becomes three again but the centers of these islands are shifted from the original ones at low eccentricity (Figure \ref{fig:f7}(f)). At even larger eccentricity, exceeding a third critical value, $e_{c3}$, another transition occurs as the original stable islands (centered at $\psi=0^\circ$, $120^\circ$ and $240^\circ$) reappear and the number of the islands becomes six again (Figure \ref{fig:f7}(g)). The values of the critical eccentricity at the transitions in the phase space structure are resolved with a finer grid in the eccentricity values (equivalently, Jacobi constant values) of the surfaces of section. For mass ratio $\mu =3 \times 10^{-5}$, the transitions occur at these critical values of the eccentricity: $e_{c1}=0.33$, $e_{c2}=0.87$ and $e_{c3}=0.91$. For other mass ratios $\mu$, the critical values of eccentricity are slightly different, but the evolution of the phase space structure is the same. The critical eccentricities for different mass ratios $\mu$ are listed in Table \ref{tab:t3} .  We note that  $e_{c1}$ and $e_{c3}$ increase whereas $e_{c2}$ decreases with increasing $\mu$; for $\mu\gtrsim(2-3)\times10^{-3}$, there is no third transition.

\begin{figure}
\figurenum{8}
\gridline{\fig{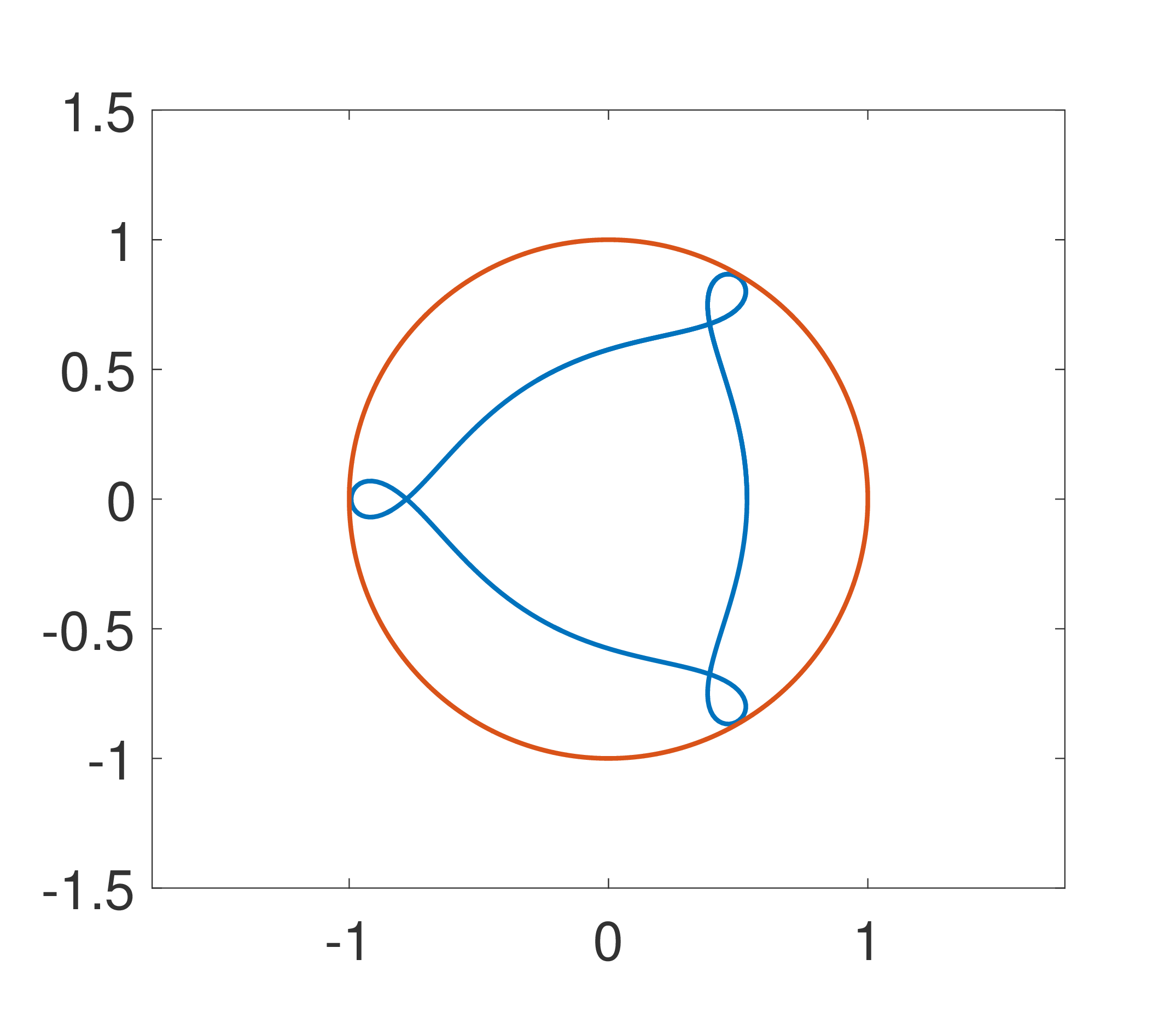}{0.3\textwidth}{(a)}
          \fig{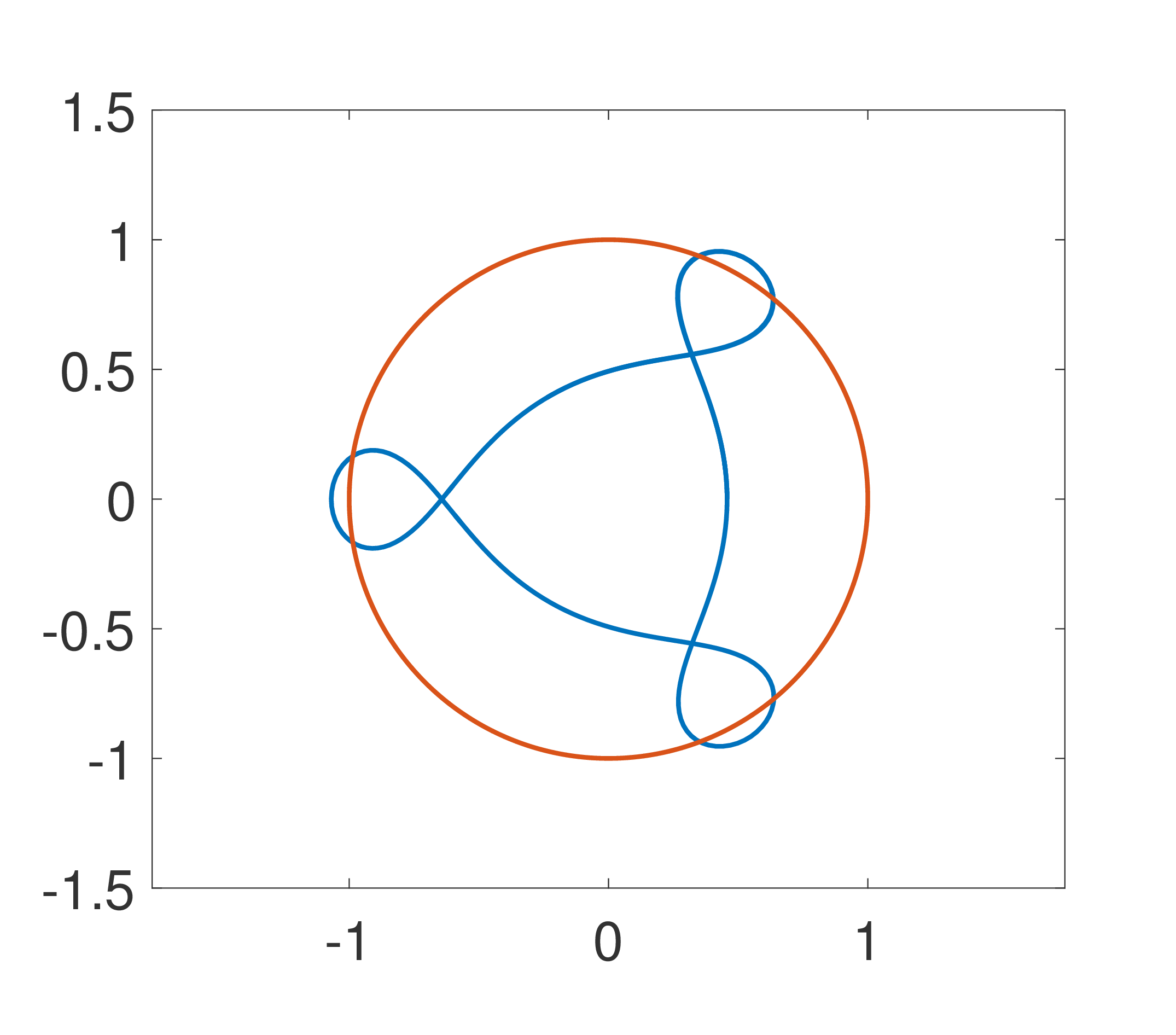}{0.3\textwidth}{(b)}
          \fig{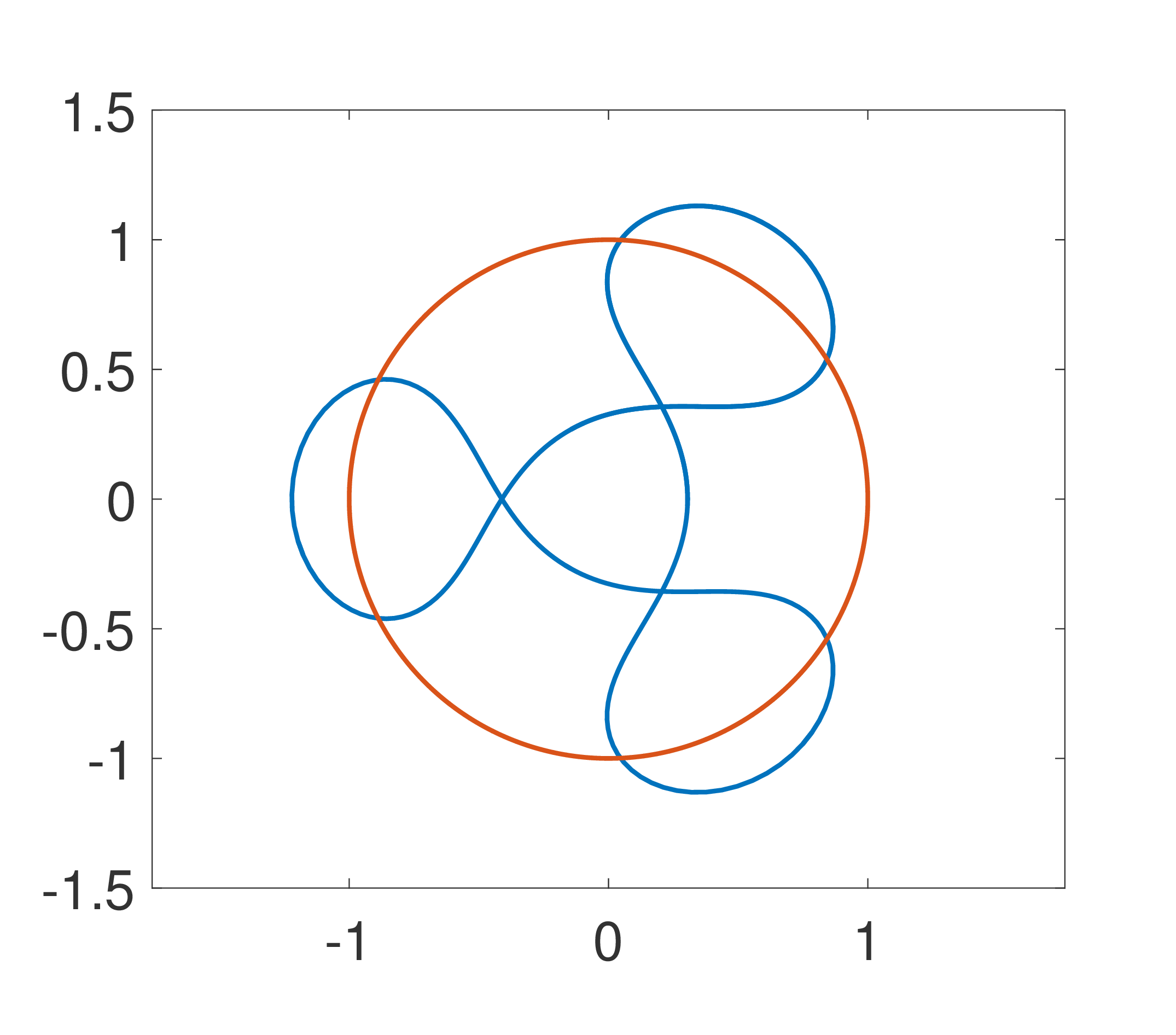}{0.3\textwidth}{(c)}
          }
\gridline{\fig{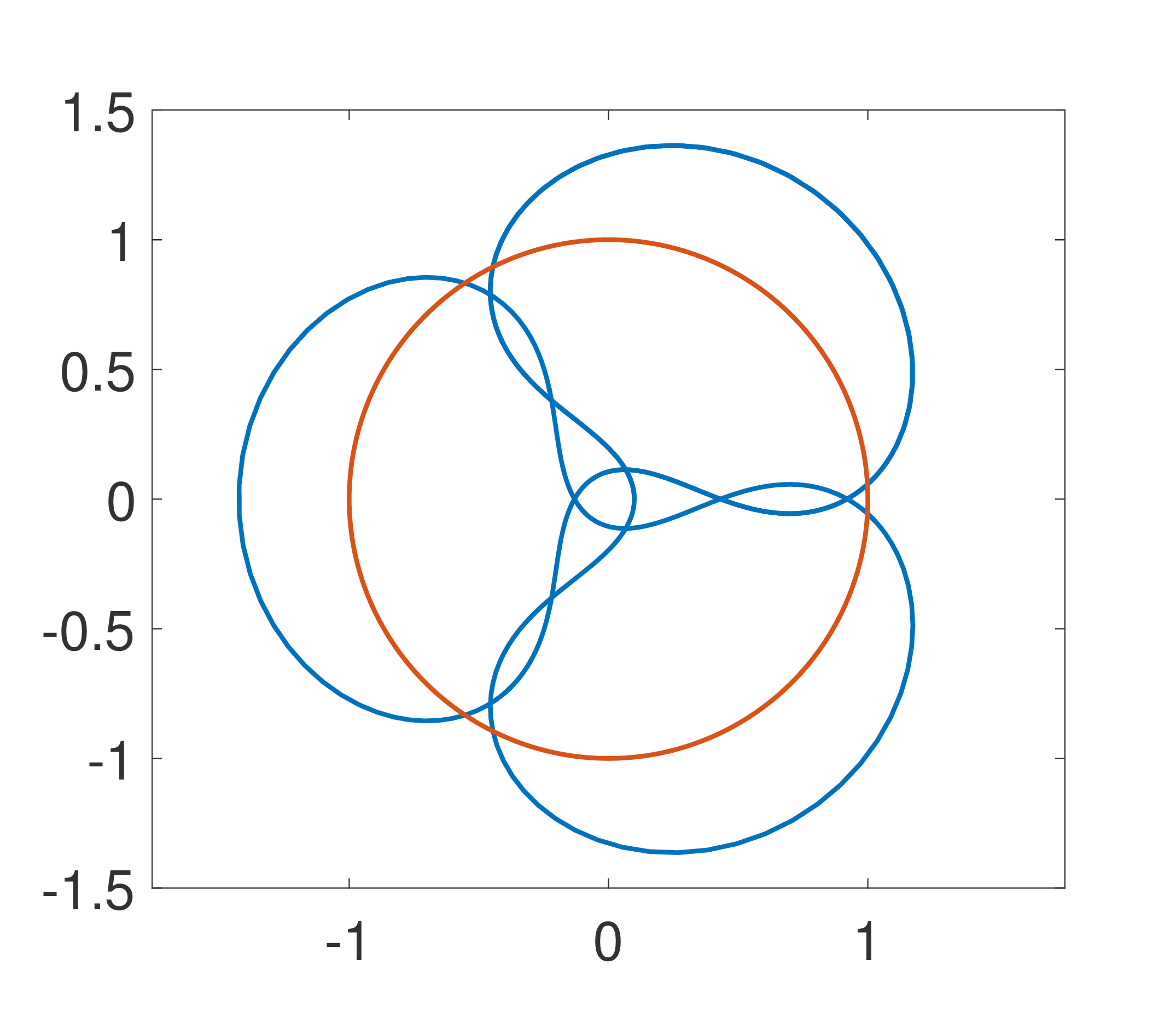}{0.3\textwidth}{(d)}
          \fig{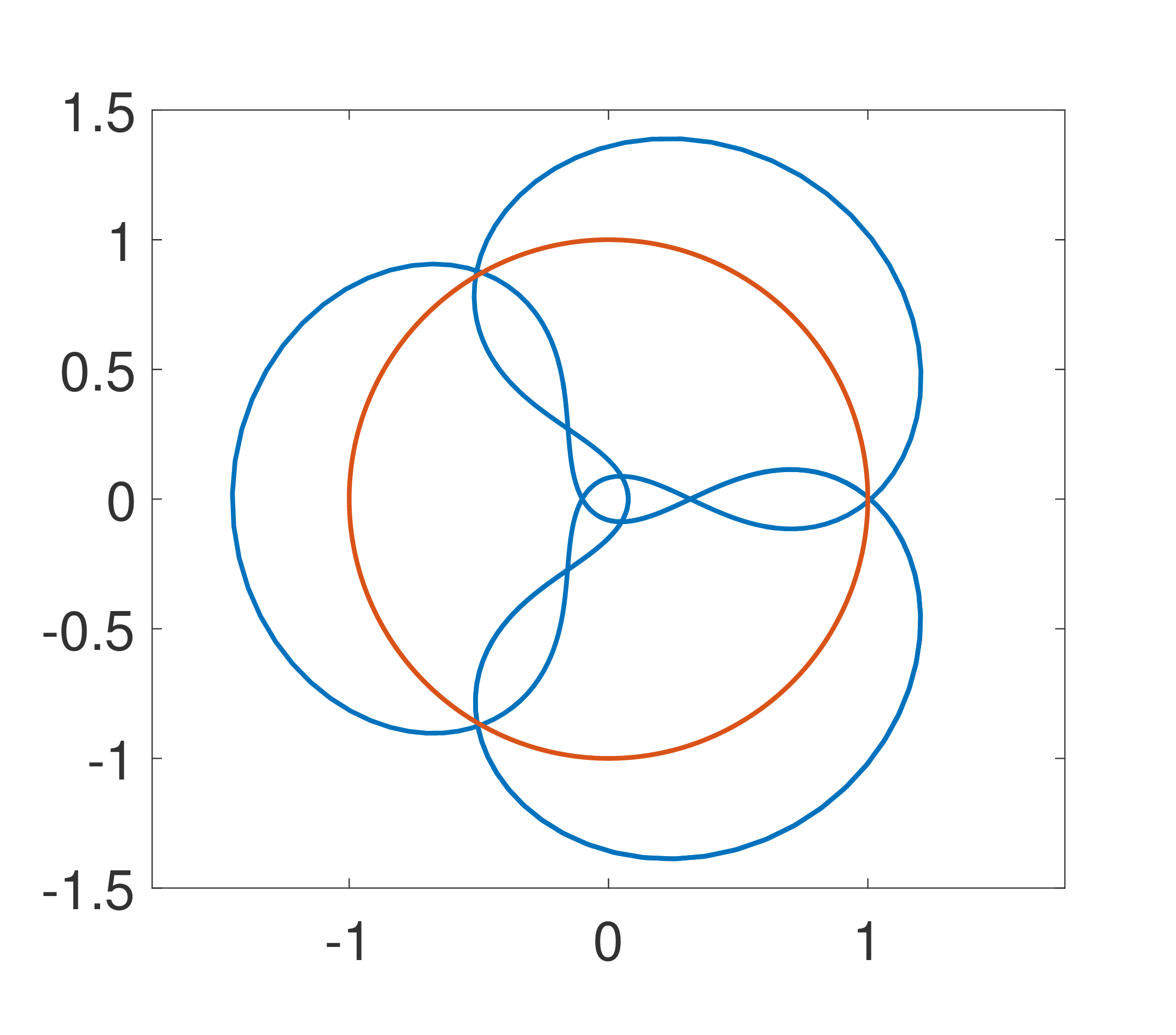}{0.3\textwidth}{(e)}
          \fig{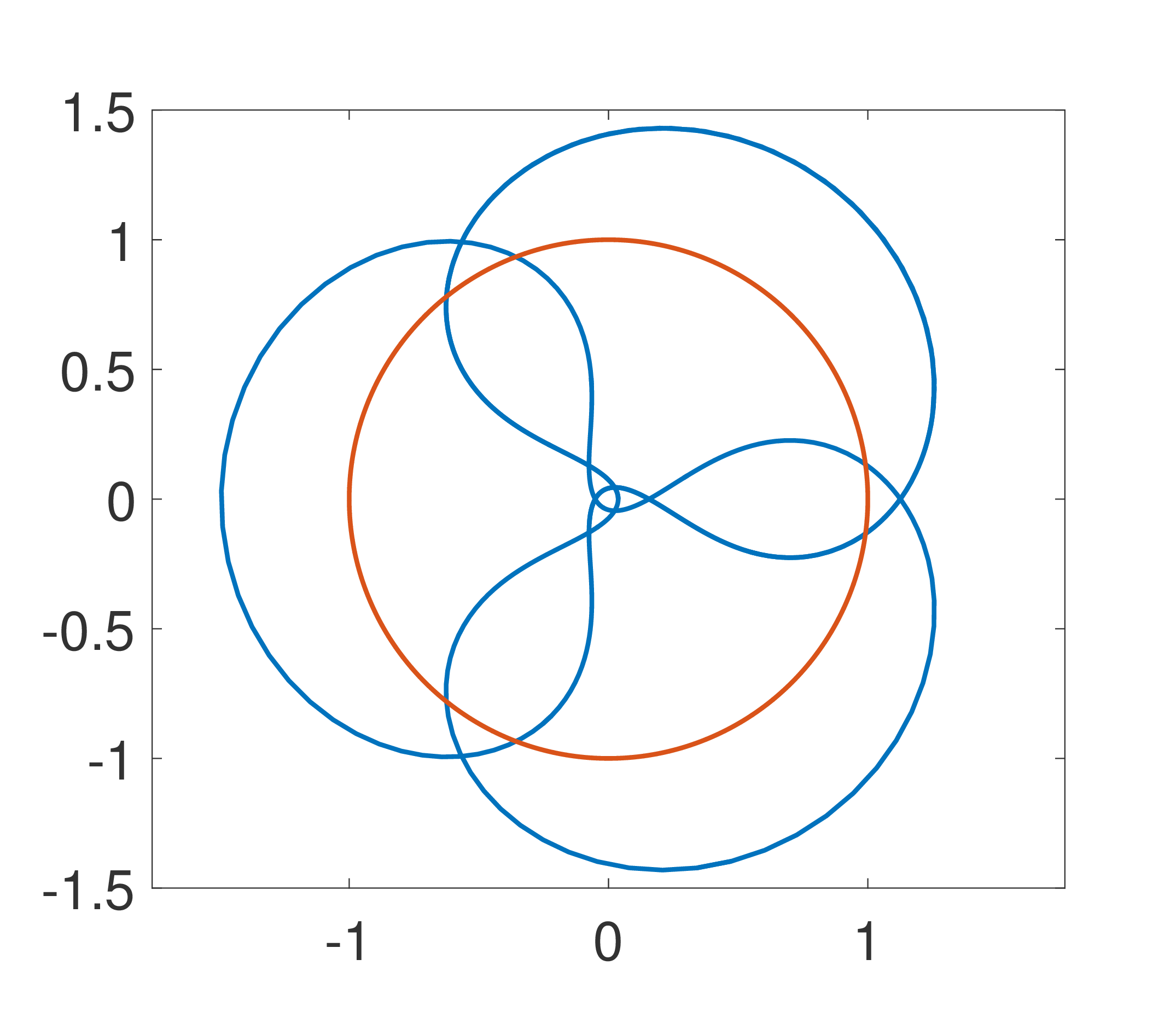}{0.3\textwidth}{(f)}
          }
\caption{Trace of the test particle's 3:2 MMR orbit in the rotating frame. The red circle is the planet's orbit of unit radius. The blue curve is the resonant orbit of the test particle in the rotating frame; there are three pericenter passages represented in this frame. From upper left to lower right, the test particle's orbit has eccentricity$e=0.30$, $e=0.40$, $e=0.60$, $e=0.87$, $e=0.90$ and $e=0.95$, respectively. For eccentricity larger than $0.31$, the test particle's aphelion exceeds the planet's orbit radius.
\label{fig:f8}}
\end{figure}

The transitions with increasing eccentricity that we find in the surfaces of section can be understood by studying the shape of the test particle's path in the rotating frame.  This is illustrated in Figure \ref{fig:f8}. When the eccentricity is small, the particle's resonant orbit is completely inside the orbit of the planet and its aphelion is far from the planet. In these cases, the perturbation of the planet is not so large, and there is no chaotic region in the surface of section. From Figure \ref{fig:f8} we can also see that the test particle's resonant orbit has a three-fold symmetry in the rotating frame, so the three centers of the stable islands in the corresponding surfaces of section have spacings of $120^\circ$ in $\psi$. For larger eccentricity, the aphelion of the test particle gets closer to the orbit of the planet, so the perturbation of the planet on the test particle also becomes larger; chaotic regions appear in the surface of section (Figure \ref{fig:f8}(a)). When the eccentricity exceeds $e_{c1}$, the test particle's orbit is planet-crossing and new stable islands appear in the surface of section. (The particle actually becomes planet-crossing when the eccentricity exceeds $0.31$, however,  for $\mu=1\times10^{-5}$, the new stable islands do not appear in the surfaces of section until eccentricity exceeds $e_{c1}=0.32$.) Once the aphelion of the test particle's orbit exceeds the orbit of the planet, the three lobes of the configuration of the particle's orbit will intersect the planet's orbit; the planet's orbit is cut into six arcs. These are where new stable islands appear in the surface of section (Figure \ref{fig:f8}(b) and \ref{fig:f7}(c)). The length of the arc of the planet's orbit which is enclosed by each of the three lobes of the test particle's orbit (in the rotating frame) limits the range of $\psi$ for the new stable islands in the surface of section, whereas the lengths of arc outside the lobes correspond to the range of $\psi$ in the old stable islands; the chaotic regions in the surface of section reflect the chaotic orbits near the intersection points where the particle would have close encounters with the planet. As the eccentricity becomes larger, the length of the arc enclosed by each lobe also becomes larger, whereas the length of arc outside the lobes shrinks. So the new stable islands grow and expand at the expense of the range of the old stable islands (Figure \ref{fig:f8}(c)). As the eccentricity approaches $0.90$, the three lobes intersect each other, and their intersection points get close to the orbit of the planet (Figure \ref{fig:f8}(d) and Figure \ref{fig:f8}(e)). At this eccentricity, the trace of the resonant orbit in the rotating frame again defines only three arcs, but these three are the ones enclosed by the lobes.  This means that the original three islands have disappeared and only the new stable islands exist. At even larger eccentricity, when the self-intersections of the lobes occur outside the orbit of the planet, as in Figure \ref{fig:f8}(f), we see that the planet's orbit is again cut into six arcs, so the ``old'' stable islands centered at $\psi=0,120^\circ,240^\circ$ reappear, albeit with smaller sizes.

\begin{figure}
\figurenum{9}
\gridline{\fig{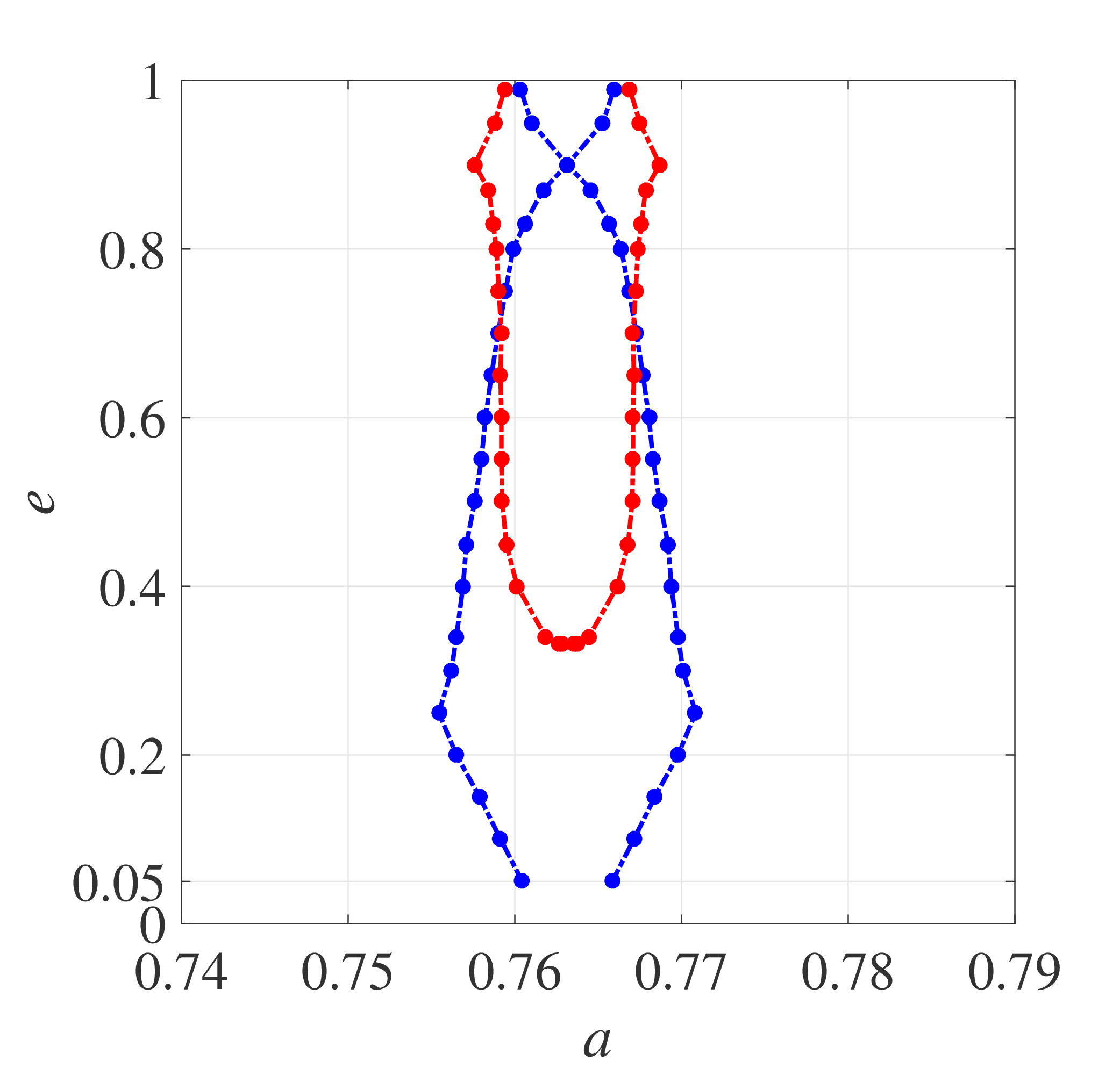}{0.3\textwidth}{(a)}
          \fig{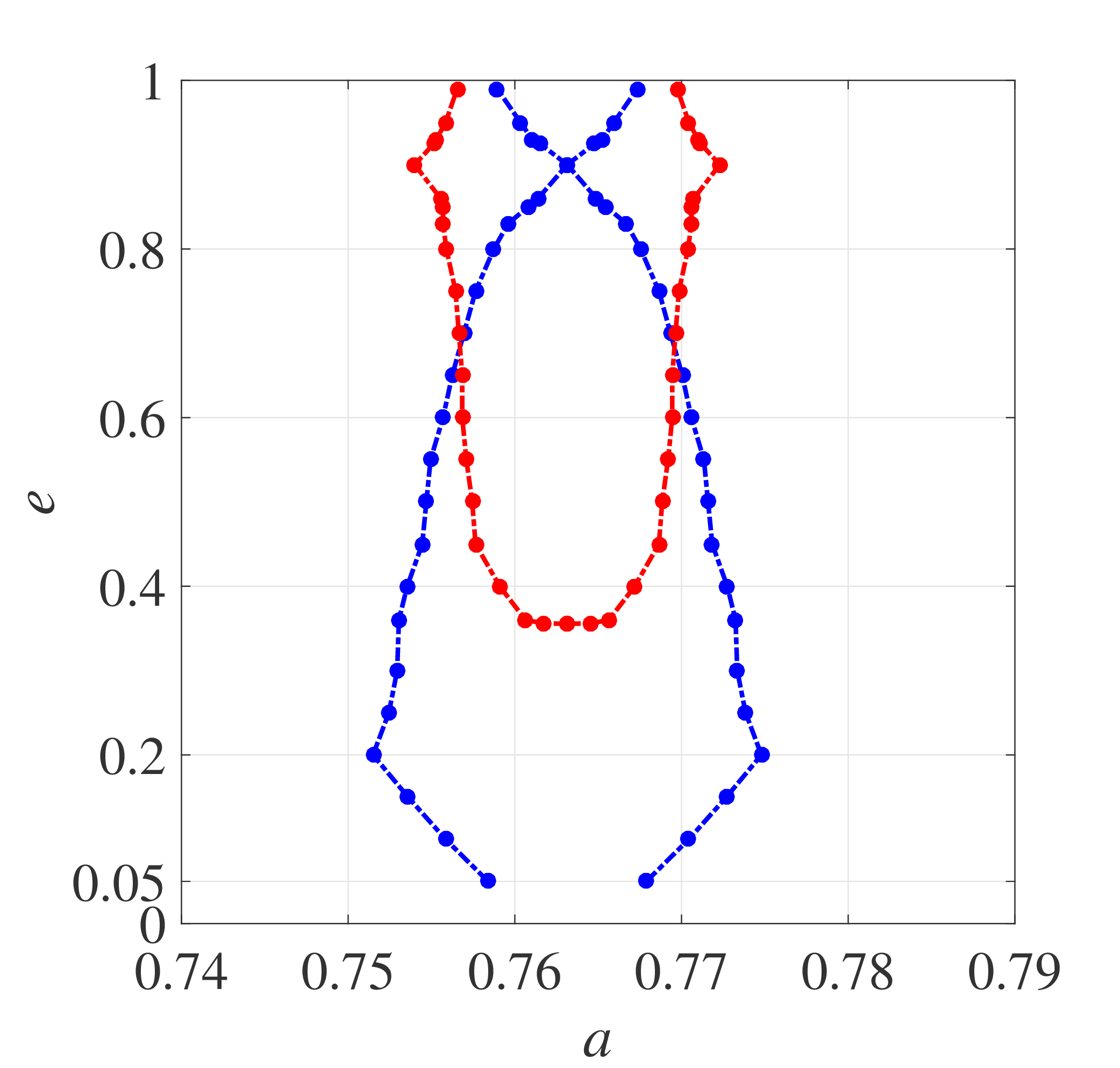}{0.3\textwidth}{(b)}
          \fig{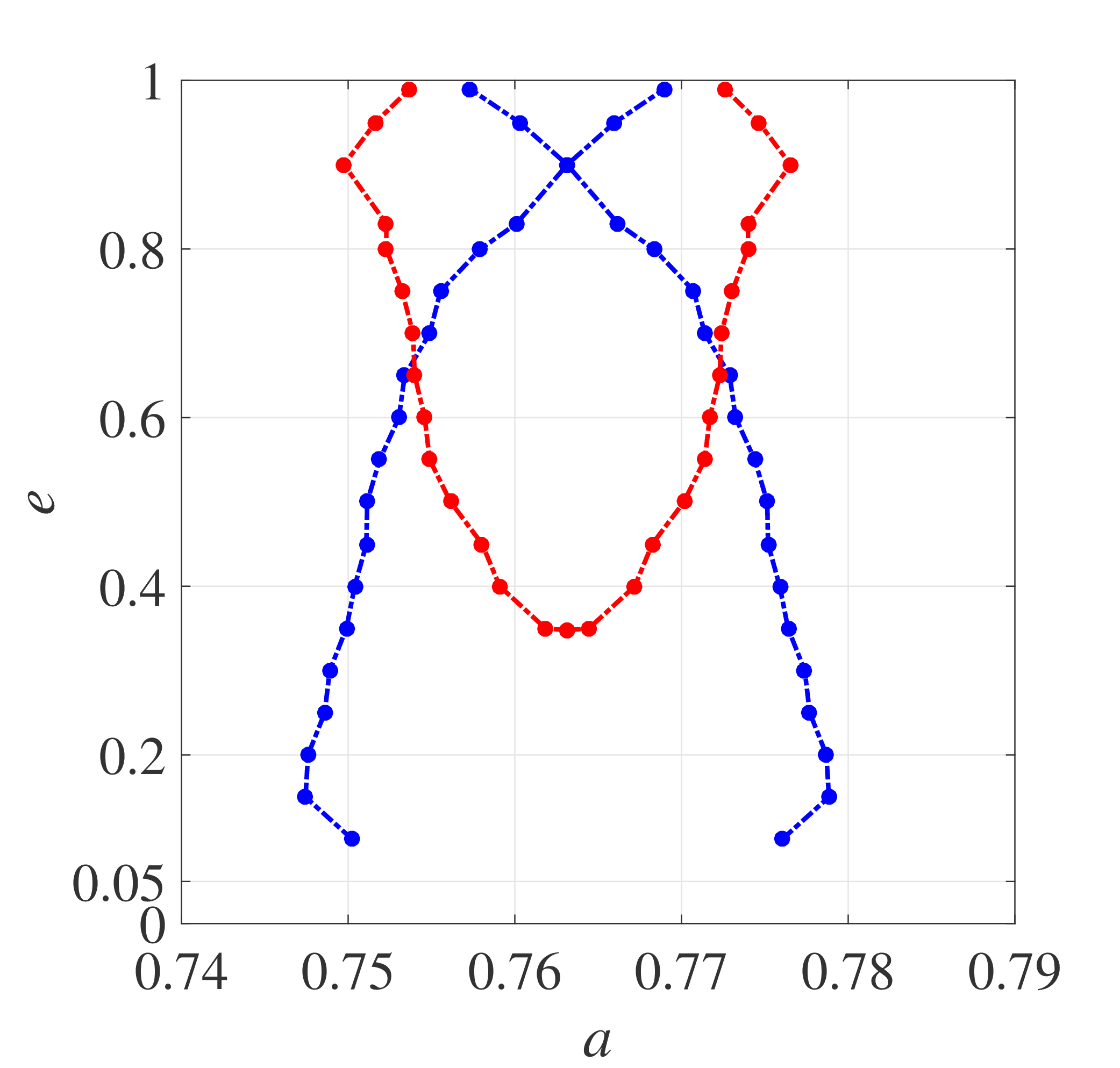}{0.3\textwidth}{(c)}
          }
\caption{Stable resonant libration zone of the interior 3:2 MMR as a function of eccentricity, for mass ratio $\mu =3 \times 10^{-5}$, $1 \times 10^{-4}$ and $3 \times 10^{-4}$, respectively. The area inside the blue line is the width in semi-major axis for the first resonance zone (with stable islands centered at $\psi=0$, $120^\circ$ and $240^\circ$). The area inside the red line is the width in semi-major axis for the second resonance zone (with stable islands centered at $\psi=60^\circ$, $180^\circ$ and $300^\circ$). \label{fig:f9}}
\end{figure}

Using the same method as for 2:1 MMR, we measured the resonance libration widths of the 3:2 MMR, for different eccentricities, $0.05\leq e\leq 0.99$ and mass ratios $\mu$.
For fixed values of $\mu$, we plot in Figure \ref{fig:f9} the stable libration regions for eccentricity ranging from $0.05$ to $0.99$. Regions inside the blue lines are the stable islands centered at $0^\circ$, $120^\circ$ and $240^\circ$ in the surface of section while regions inside the red lines are stable islands centered at $60^\circ$, $180^\circ$ and $300^\circ$. We refer to the former as the ``first resonance zone" and the latter as the ``second resonance zone". For small eccentricities, the first resonance zone increases in width with increasing eccentricity, reaches a maximum width, then decreases in width until it vanishes near $e= e_{c2}$, and then begins to increase again. Once it appears near $e= e_{c1}$, the second resonance zone width increases, and reaches maximum width when the first resonance zone vanishes near $e_{c2}$, and then decreases. From Figure \ref{fig:f9} we can see that the first resonance zone reaches its maximum width just before the second resonance zone appears. Table \ref{tab:t3} lists the values of the eccentricity, $e_m$, where the first resonance zone has its maximum width, for each mass ratio $\mu$ that we investigated. We find that $e_m$ decreases with increasing $\mu$, i.e., the maximum width occurs at smaller eccentricity for larger $\mu$.  The maximum of the resonance width is monotonically larger for larger mass ratio $\mu$. And the second resonance zone reaches its maximum width when the eccentricity is around $e=0.90$, which is also when the width of the first resonance zone is vanishingly small.

\floattable
\begin{deluxetable}{ccccc}
\tablecaption{Values of $e_m$ and $e_{c1},e_{c2},e_{c3}$ for the interior 3:2 MMR  \label{tab:t3}}
\tablehead{
\colhead{Mass ratio, $\mu$} & \colhead{$e_m$} & \colhead{$e_{c1}$} & \colhead{$e_{c2}$}& \colhead{$e_{c3}$}
}
\startdata
  $1 \times 10^{-5}$     & $0.28$   & 0.320 & 0.885 & 0.906 \\
  $3 \times 10^{-5}$     & $0.25$   & 0.332 & 0.877 & 0.912 \\
  $1 \times 10^{-4}$     & $0.22$   & 0.356 & 0.862 & 0.925 \\
  $3 \times 10^{-4}$     & $0.18$   & 0.395 & 0.835 & 0.945 \\
  $1 \times 10^{-3}$     & $0.15$   & 0.465 & 0.777 & 0.958 \\
  $3 \times 10^{-3}$     & $0.14$   & 0.551 & 0.741 & -- \\
\enddata
\tablecomments{$e_m$ represents the eccentricity where the first resonance zone has its maximum width, $\Delta a$,
and $e_{c1},e_{c2},e_{c3}$ are the transition eccentricities when the stable islands in phase space change structure.
}
\end{deluxetable}

\begin{figure}
\figurenum{10}
\gridline{\fig{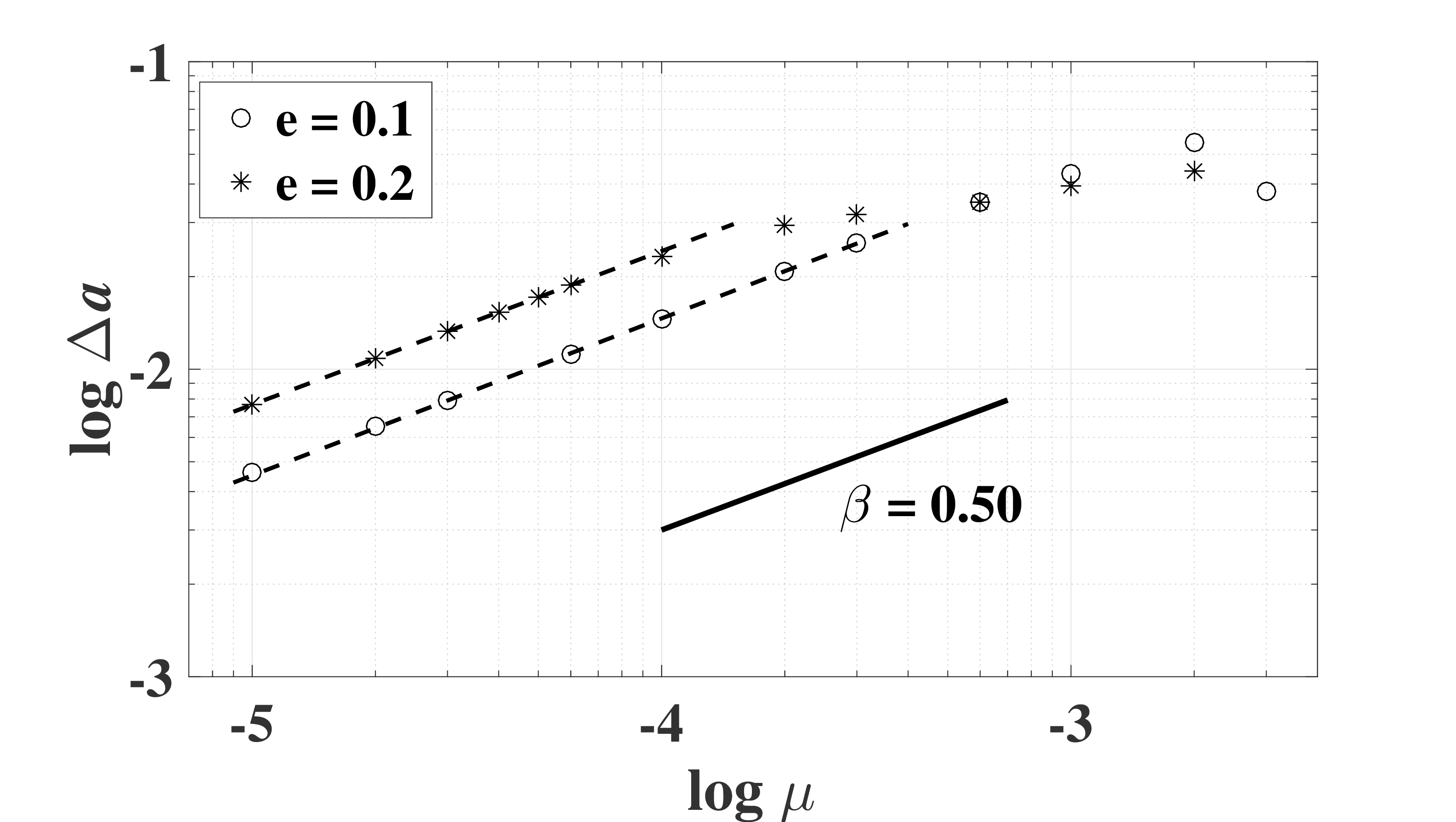}{0.5\textwidth}{(a)}
          \fig{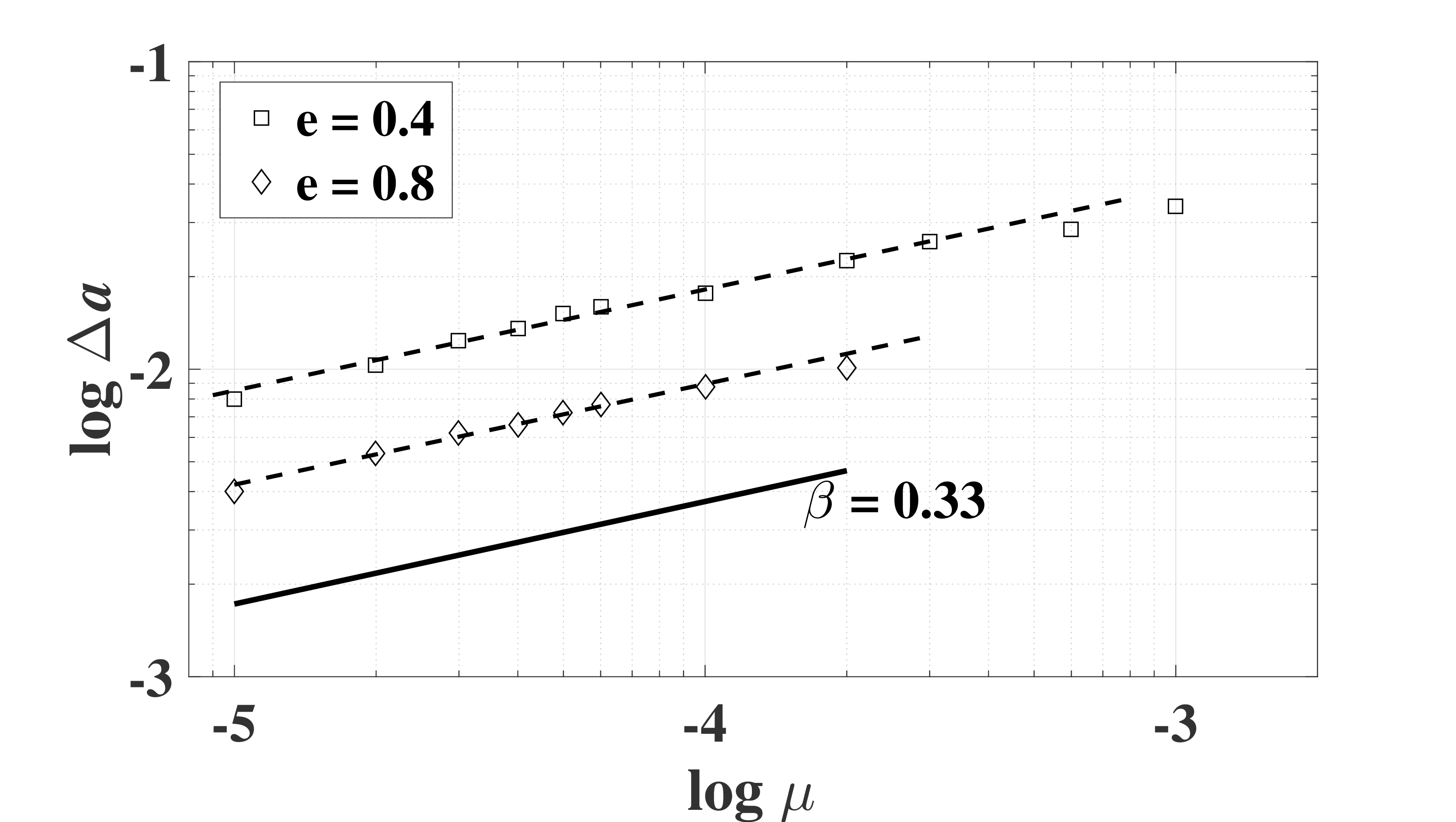}{0.5\textwidth}{(b)}
          }
\caption{Planet-mass dependence of the width, $\Delta a$, of the 3:2 MMR's ``first resonance zone": (a) for low eccentricities, $e=0.1$ (circles), $e=0.2$ (stars), and (b) for higher eccentricities, $e=0.4$ (squares), $e=0.8$ (diamonds). The dashed lines are the best-fit power-law relation between $\Delta a$ and mass ratio $\mu$.  The black solid line is a reference line with the indicated slope value. \label{fig:f10}}
\end{figure}

As before, we postulate that, for fixed eccentricity, the resonance width $\Delta a$ has a power law relation with the mass of the planet (Equation \eqref{eq23}). (Similar to the case of the 2:1 MMR, the dependence on the eccentricity is not described well as a power law, and we have not attempted to fit a functional form for it.)
The results for the first resonance zone are shown in Figure \ref{fig:f10}. For $e=0.1$ and $e=0.2$, we find that the best-fit power-law relations are $\Delta a=1.611\mu^{0.5105}$ and $\Delta a=2.410\mu^{0.4995}$, respectively, which shows that the power-law index is near $1/2$ (Figure \ref{fig:f10}(a)).
For larger eccentricity, $e=0.4$ and $e=0.8$, the best-fit power-law relations are $\Delta a=0.3746\mu^{0.3286}$ and $\Delta a=0.1822\mu^{0.3272}$, respectively, which shows that the power-law index is near $1/3$ (Figure \ref{fig:f10}(b)), different from that at smaller eccentricity.  It is worth noting that for these larger eccentricities, the resonance width $\Delta a$  becomes smaller as the eccentricity increases.

\begin{figure}
\figurenum{11}
\plotone{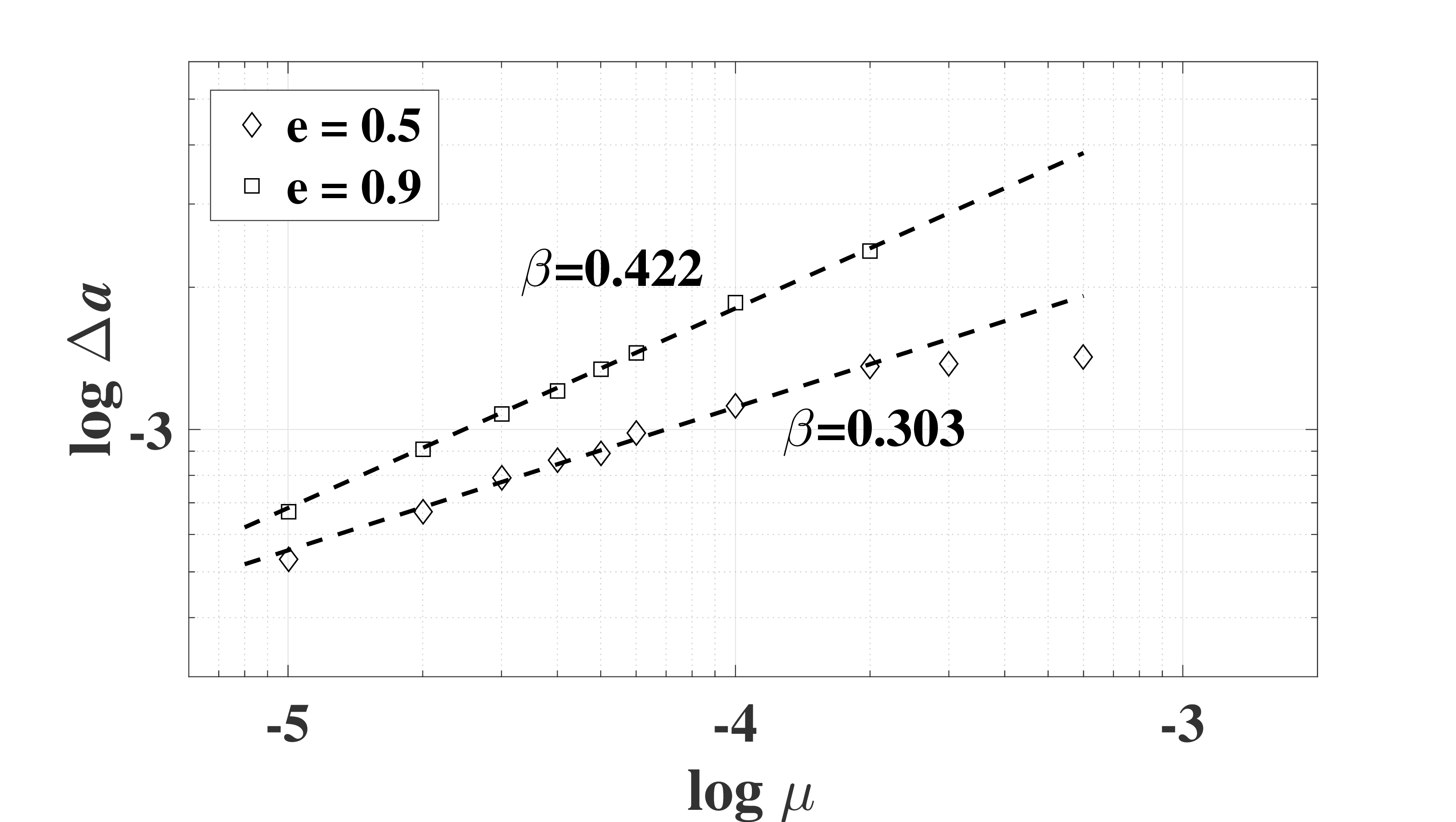}
\caption{Planet-mass dependence of the width, $\Delta a$, of the 3:2 MMR's ``second resonance zone", for eccentricity $e=0.5$ (diamonds) and $e=0.9$ (squares). The dashed lines are the best-fit power-law relation between $\Delta a$ and mass ratio $\mu$.
\label{fig:f11}}
\end{figure}

For the second resonance zone, Figure \ref{fig:f11} shows the results for the power-law relation for $e=0.5$ and $e=0.9$; the best-fit power-laws relations are found to be $\Delta a=0.1811\mu^{0.3027}$ and $\Delta a=0.8821\mu^{0.4223}$, respectively. The power-law index is different for different eccentricities, larger for larger eccentricity.

We notice that for both resonance zones, there is a rollover in the slope of the $\log\Delta a$--$\log\mu$ curve at larger $\mu$; we can see this drop-off in the tail of the data in Figure \ref{fig:f10}--\ref{fig:f11}.  The reason for the rollover appears to be that when the mass ratio is big enough, the chaotic regions around the stable islands limit the growth of the stable libration region. In computing the reported best-fit power-law fits, we excluded the data for the large mass ratios, $\mu\gtrsim(2-3)\times10^{-4}$.

Table \ref{tab:t4} summarizes the results for the best-fit power-law relations for the first and second resonance zones for different eccentricities.

\begin{deluxetable}{ccccc}
\tablecaption{Best-fit power-law parameters of the width of the first and second resonance zones of the interior 3:2 MMR \label{tab:t4}}
\tablehead{
\colhead{} & \colhead{First zone} & \colhead{} & \colhead{Second zone} & \colhead{} \\
\colhead{Eccentricity} & \colhead{$\alpha$} & \colhead{$\beta$} & \colhead{$\alpha$} & \colhead{$\beta$}
}
\startdata
0.1 & 1.611           & 0.5105           & --          & -- \\
0.2 & 2.410           & 0.4995  & --          & -- \\
0.3 & 0.3124          & 0.2970  & --          & -- \\
0.4 & 0.3746          & 0.3286           & 0.05749    & 0.2107 \\
0.5 & 0.3053          & 0.3168           & 0.1811     & 0.3027     \\
0.6 & 0.2853          & 0.3263          & 0.2155     & 0.3154     \\
0.7 & 0.2753          & 0.3368           & 0.2723     & 0.3346     \\
0.8 & 0.1822          & 0.3272           & 0.3885     & 0.3650     \\
0.9 & --          & --           & 0.8821     & 0.4223     \\
\enddata
\tablecomments{Eccentricities less than 0.4 are not shown here because the second resonance zone doesn't appear until $e \gtrsim (0.3-0.4)$. Similarly, the first resonance zone doesn't exist when $e \sim 0.9$.
}
\end{deluxetable}

\section{Summary and discussion} \label{sec:conclusion}

In most previous studies, the widths of mean motion resonances have been investigated in the low eccentricity regime with perturbative analytical methods and it has been commonly assumed that the stable resonance widths vanish as MMRs become unstable and chaotic at high eccentricities owing to resonance overlap \citep{Murray99}.
In the present study, we carried out non-perturbative numerical analyses of MMRs in the circular planar restricted three body problem, for a wide range of eccentricities $e$ and mass ratios $\mu$.  A surface of section is introduced to study the phase space structure near resonances. We located the resonance libration centers and computed the widths of stable zones of the interior 2:1 and 3:2  MMRs. Our investigation shows that stable MMR zones change with increasing eccentricity, with one or more transitions of the dominant stable islands, but sizable stable libration zones still exist at high eccentricities.

For the case of the interior 2:1 MMR, when the eccentricity is smaller than the planet-crossing value, the surface of section shows a chain of two stable islands centered at $\psi=0^\circ$ and $180^\circ$, where $\psi$ is the angular separation of the planet from the test particle at pericenter.
With increasing eccentricity, the width, $\Delta a$, of the stable libration zones expands even as chaotic regions appear near the resonance separatrix.
The stable resonance width reaches a maximum and then decreases with increasing eccentricity; the maximum width occurs at eccentricity values a little below the planet-crossing eccentricity.  A new chain of two stable islands centered at $\psi=90^\circ,270^\circ$ appears when the eccentricity reaches a critical value, which is close to but somewhat larger than the planet-crossing eccentricity.
The widths of the new libration islands grow while the old ones shrink with increasing eccentricity beyond the critical value. The critical eccentricities for different values of $\mu$ are listed in Table~\ref{tab:t1}.

For the case of the interior 3:2 MMR, the centers of the stable libration zones are located at $\psi=0^\circ, 120^\circ$, and $240^\circ$ when the eccentricity is small.
With increasing eccentricity, the stable islands expand and the separatrix becomes a chaotic layer.  When the eccentricity exceeds a critical value, a new chain of three islands appears, centered at $\psi=60^\circ, 180^\circ$, and $300^\circ$.
The new stable islands grow and expand while the old ones shrink with increasing eccentricity. The three old stable zones vanish at a second critical value of the eccentricity, but they reappear and grow again when the eccentricity becomes larger. So, with increasing eccentricity, there are three transitions in the phase space structure of the 3:2 MMR. The critical eccentricities for different values of $\mu$ are listed in Table~\ref{tab:t3}.

We have shown that these transitions are related mainly to the geometry of the trace of the particle's resonant orbit in the rotating frame; they depend only weakly on the mass ratio $\mu$. For eccentricity below the first critical transition, the chain of stable islands in the surfaces of section corresponds to stable resonant librations about zero of the usual resonant angle, $\phi=2\lambda'-\lambda-\varpi$ and $\phi=3\lambda'-2\lambda-\varpi$ for the 2:1 and 3:2 MMRs, respectively.
Physically, these correspond to conjunctions of the planet and test particle when the particle is near pericenter; they can be called ``pericentric librations".
The first transition occurs when the particle's orbit becomes a planet-crossing orbit and a new chain of stable islands appears in the phase space; these correspond to librations of the resonant angle $\phi$ about $180^\circ$. Physically, these are stable librations of conjunctions of the planet and test particle when the particle is near apocenter; they can be called ``apocentric librations". To our knowledge, this stable ``apocentric libration" in the interior MMRs at high eccentricity has not been reported in previous studies. Additionally, for the 3:2 MMR case, we have shown that a second and a third transition occurs when the self-intersections of the trace of the test particle's resonant orbit in the rotating frame occur at heliocentric distance exceeding the orbit radius of the planet. Each of these transitions makes the number of the stable islands halve or double, as can be seen in Figure \ref{fig:f2} and Figure \ref{fig:f7}.  We showed how these transitions can be understood physically by relating to the geometry of the eccentric resonant orbits in the rotating frame (Figure~\ref{fig:f3} and Figure~\ref{fig:f8}).

The non-perturbative numerical analyses shows that high eccentricity MMRs have nearly as large stable libration zones as low eccentricity MMRs. We measured the resonant widths of the libration zones in the surfaces of section. Examples are plotted in Figure \ref{fig:f4} and Figure \ref{fig:f9}. We fit an empirical power-law relation between the resonant widths and the mass of the planet, $\Delta a = \alpha\mu^\beta$. For the first resonance zone (pericentric librations) of the 2:1 MMR, the power law index $\beta$ is close to $0.50$ and $0.40$, respectively, for eccentricities smaller and larger than $e_m$, where $e_m$ is the value of eccentricity at maximum resonance width.
For the first resonance zone (pericentric librations) of the 3:2 MMR, we find that $\beta$ is also close to $0.50$ for eccentricities below $e_m$, but for higher eccentricity, it is close to $0.33$. (The values of $e_m$ depend upon the mass ratio $\mu$ and are listed in Table~\ref{tab:t1} and Table~\ref{tab:t3}.)
The second resonance zone (apocentric librations) exists only for eccentricities exceeding the planet-crossing value; we find that $\alpha$ and $\beta$ both increase monotonically with increasing eccentricity; the results are summarized in Table~\ref{tab:t2} and Table~\ref{tab:t4}.

\cite{Murray99} provide a plot of the resonance libration widths, $\Delta a$, of a few interior MMRs of Jupiter, using a perturbative analysis of the circular planar restricted three body model (see their Figure 8.7). Their calculation is restricted to $\mu=9.55\times10^{-4}$ and to the eccentricity range of 0 to 0.3 (pertinent to the Sun--Jupiter--asteroid problem). Our results for the interior 2:1 and 3:2 MMRs for $\mu=10^{-3}$ agree fairly well with these analytical results over the eccentricity range 0.05--0.3, although the analytical theory somewhat overestimates the resonance width of the 3:2 MMR (possibly because it does not account for the chaotic boundary).  More generally, we note that the results of analytical perturbation theory predict that the resonance widths (of first order resonances such as the 2:1 and 3:2) increase with particle eccentricity and planet mass as $\Delta a \sim (e\mu)^{1/2}$~\citep{Murray99}. Our non-perturbative numerical results confirm this behavior for small $\mu$ and small eccentricities, $\mu\lesssim10^{-5}$ and $0.05\lesssim e\lesssim0.1$.  For larger $\mu$ and eccentricities up to $e_m$, the mass dependence is as expected, $\Delta a \sim \mu^{1/2}$, but the eccentricity dependence is more complicated.  For larger eccentricities, $e>e_m$, the resonance widths of the pericentric librations decrease with increasing eccentricity and have weaker mass dependence than the theoretical prediction.  The low order perturbative analysis misses entirely the transitions in the resonance phase space and the appearance of the apocentric libration zone at higher eccentricities.

In the present work, we have investigated only the 2:1 and 3:2 interior MMRs.  In a future study, it would be of interest to extend this investigation to other MMRs.

\bigskip

X. W. acknowledges funding from National Basic Research Program of China (973 Program) (2012CB720000) and China Scholarship Council. R. M. acknowledges funding from NASA (grant NNX14AG93G) and NSF (grant AST-1312498).



\end{document}